\documentclass[11pt,a4paper]{article}

\usepackage[T1]{fontenc}
\usepackage{graphicx}
\usepackage{longtable}
\usepackage{braket}
\usepackage{tabularx}
\usepackage{enumerate}
\usepackage{url}
\usepackage{amsmath}
\usepackage{subcaption}
\usepackage{float}
\usepackage{tikz}
\usepackage{ragged2e}
\usepackage{arydshln}

\usetikzlibrary{arrows.meta,positioning,calc}

\definecolor{flowcol}{RGB}{0,105,92}

\begin{document}

\title{Machine Learning Assisted Parametrisation and Prediction of 
Bare to Neutral $\beta^-$ Decay Rate Ratios of Fully Ionised Atoms}

\author{
Arkabrata Gupta$^{1,2,\dagger}$
\and
Spandan Aich$^{2,3, \dagger}$
\and
Suparna Sau$^{1,2}$
\and
Sangeeta Das$^{1}$
}

\date{}

\maketitle

\begin{center}
\small
$^{1}$Department of Basic Science and Humanities, 
Institute of Engineering \& Management, 
University of Engineering and Management, Kolkata, 700160, India\\

$^{2}$Centre of Excellence in Astronomical Studies, 
Institute of Engineering \& Management, 
University of Engineering and Management, Kolkata, 700160, India\\

$^{3}$Department of Computer Science Engineering, 
Institute of Engineering \& Management, 
University of Engineering and Management, Kolkata, 700160, India\\

\vspace{0.5em}

$^{*}$Corresponding author: \texttt{arkabratagupta@gmail.com}\\
$^{\dagger}$These authors contributed equally.
\end{center}


\begin{abstract}
\justifying
The $\beta^-$ decay scenario of a nucleus differs significantly from its terrestrial characteristics when the atom is highly ionised or fully stripped of its electrons. Under such conditions, in addition to the conventional $\beta^-$ decay to atomic continuum, the emitted electron may occupy a vacant atomic orbital of the daughter atom, giving rise to bound state $\beta^-$ decay. Such environments occur naturally in stellar interiors and can also be produced in storage ring or plasma trap experiments. The relative contributions of the continuum and bound state decay channels depend on several nuclear and atomic properties, such as the decay $Q$ value, the mass and proton numbers of the daughter nucleus. Consequently,  predicting decay rate enhancements becomes a complex problem that generally requires detailed theoretical calculations. In this work, we explore the use of machine learning (ML) techniques to investigate the systematics of $\beta^-$ decay in fully ionised atoms. ML based regression models are developed using theoretically calculated decay rate data for a set of astrophysically relevant allowed $\beta^-$ transitions. A global parametric expression is proposed to estimate the bare-to-neutral decay rate ratio, and its predictive capability is examined using independent validation data. In addition, Random Forest and Artificial Neural Network models are trained to predict the bare-to-neutral decay rate ratio directly from a set of nuclear and atomic parameters. The results show that ML methods can successfully capture the underlying trends governing decay rate enhancement and provide rapid estimates without computationally demanding calculations. The developed models offer a practical tool for estimating the maximum possible $\beta^-$ decay rates and will be useful for applications in nuclear astrophysics, future storage ring or plasma trap experiments, and nucleosynthesis modeling.
\end{abstract}

\section{\label{introduction} Introduction}

\justifying

$\beta^-$ decay is a weak interaction process in which a neutron inside an atomic nucleus transforms into a proton with the emission of an electron and an antineutrino into the continuum. In partially or fully ionised atoms, however, continuum state $\beta^-$ decay is not the only possible decay channel. Under such conditions, the emitted electron may also be captured in a vacant atomic orbital of the daughter atom, giving rise to bound state $\beta^-$ decay. The concept of bound state $\beta^-$ decay was first proposed by Daudel \textit{et al.} in 1947 \cite{Daudel1947} as the time-reversed analogue of orbital electron capture.

For neutral atoms, the atomic orbitals are already occupied, leaving essentially no available phase space for the emitted $\beta^-$ electron to be captured into a bound state. Consequently, the contribution of bound state decay is zero or negligible. In contrast, highly ionised or bare atoms possess vacant atomic orbitals, allowing the emitted electron to occupy bound electronic states along with the possibility to go into continuum. The resulting enhancement of the decay phase space can significantly modify the total $\beta^-$ decay rate. Highly ionised atoms naturally occur in stellar environments, where extreme temperatures and densities can strip atoms of most or all of their electrons \cite{TakahashiBoyd1987,  Gupta2019}. In particular, many heavy nuclei can exist in fully ionised states in stellar interiors. Similar conditions can also be realized terrestrially in storage ring experiments, where ions are accelerated and stripped of their bound electrons.

Further theoretical investigations of bound state $\beta^-$ decay was performed in the 1960s and 1970s, primarily focusing on low-$Z$ nuclei such as tritium \cite{Bahcall1961}. Many of these early studies relied on limited experimental data and incomplete theoretical treatments. A major advance was achieved by Takahashi and Yokoi, who performed systematic studies of $\beta^-$ decay in highly ionised stellar environments \cite{Takahashi1983,Takahashi1987}. Although these works did not explicitly separate the bound state contribution, a subsequent study by Takahashi and Boyd \cite{TakahashiBoyd1987} provided bound state decay rates for several important $s$-process nuclei. Collectively, these investigations demonstrated that the inclusion of bound state decay can substantially change/ enhance the decay rates of highly/ fully ionised atoms.

Experimental confirmation of this phenomenon was first achieved by Jung \textit{et al.} in 1992 through the observation of bound state $\beta^-$ decay in fully ionised $^{163}$Dy ions \cite{Jung1992}. Later, measurements were performed for $^{187}$Re, $^{207}$Tl, and $^{205}$Tl using storage ring facilities \cite{Bosch1996, Ohtsubo2005, Leckenby2024}. More recently, the PANDORA project has been developed to reproduce stellar plasma like conditions in the laboratory, enabling investigations of $\beta^-$ decay in ions with different charge states under controlled plasma environments \cite{mishra2022, Mascali2022}.

Despite these experimental advances, relatively limited theoretical development occurred for nearly three decades. In 2019, Gupta \textit{et al.} revisited the problem by calculating $\beta^-$ decay rates of bare nuclei in the mass range $A=60$--240 using updated nuclear and atomic data \cite{Gupta2019}. These calculations established upper limits for bound state $\beta^-$ decay rates under terrestrial conditions. In a subsequent study, Gupta \textit{et al.} investigated the contribution of bound state decay to stellar $s$-process environments for nuclei in the mass region $A=59$--81 \cite{Gupta2023}. In contemporary studies, Liu \textit{et al.} \cite{Liu2021,Liu2022} and Mishra \textit{et al.} \cite{Mishra2022b, Mishra2024} also evaluated bound state $\beta^-$ half-lives of bare atoms. Furthermore, the ionisation induced reduction of half-lives in long-lived radionuclides has attracted attention because of its potential implications for nuclear waste transmutation and management, particularly for long-lived fission products \cite{Gupta2025}.

For highly ionised atoms, the total $\beta^-$ decay rate is the sum of the continuum state and bound state contributions. The competition between these two channels depends on several interconnected nuclear and atomic properties, including the transition $Q$ value, the degree of ionisation, ionisation potential depression, and the atomic and mass numbers of the daughter nucleus. The strong coupling among these parameters leads to considerable complexity, making reliable predictions difficult within purely analytical or semi analytical frameworks. Consequently, detailed theoretical calculations are generally required to determine the decay properties of highly ionised atoms under specific physical conditions \cite{Gupta2019, Takahashi1983, Takahashi1987, Gupta2023}.

In this context, artificial intelligence and machine learning (AIML) techniques provide a promising alternative approach. Machine learning (ML) algorithms are particularly effective in identifying complex nonlinear correlations and hidden patterns in multidimensional data sets without requiring explicit analytical functional forms \cite{Carleo2019}. Such capabilities are especially valuable in the present concept of bound state $\beta^-$ decay, where experimental information is often limited and theoretical descriptions involve many body interactions and competing physical effects. In recent years, ML methods have been successfully applied to a broad range of nuclear physics problems, including nuclear mass predictions, decay property estimations, uncertainty quantification, parameter optimization, and the extraction of interpretable physical models \cite{Boehnlein2022,Lu2025,Shree2026}.

In the present work, we explore ML based GPR and SR techniques for predicting the enhancement of $\beta^-$ decay rates in fully ionised atoms. Since fully ionised atoms represent the limiting case of maximum ionisation, the corresponding decay rates provide the maximum possible enhancement relative to neutral atoms. The present study is restricted to allowed $\beta^-$ transitions. Using our own theoretically calculated decay rates as training data, we develop both empirical and machine learning based predictive models. Our primary objective is to derive a global parametric expression capable of reproducing the bare-to-neutral decay rate ratio without requiring detailed theoretical calculations. In parallel, Random Forest (RF) and Artificial Neural Network (ANN) models are trained to predict the same quantity directly from relevant nuclear and atomic parameters. Through these approaches, we investigate the systematic dependence of decay rate enhancement on the underlying physical variables and assess the capability of ML techniques to uncover trends that may not be readily evident within conventional theoretical frameworks. Finally, the developed models are applied to a large set of nuclei mentioned by Takahashi and Yokoi \cite{Takahashi1987}, for which bare atom decay rates are presently unavailable.

The paper is organized as follows. Section~\ref{methodology} presents the theoretical framework for calculating bare atom $\beta^-$ decay rates together with the machine learning methodology employed in this work. The data sources and preparation procedures are described in the subsequent section \ref{data sources}. The results and discussion, section \ref{results} provide details of the development of the parametric model and its validation, followed by the performance of the RF and ANN models. A comparative assessment of all three approaches is then presented. The developed models are subsequently used to predict decay rate ratios for a large number of nuclei for which such data are not currently available. Finally, the main conclusions are summarized in the concluding section \ref{summary}.


\section{\label{methodology}Methodology}

\subsection{\label{theory} Theoretical Framework}

In this work, we have considered only the allowed $\beta^-$ transitions. The transition rates are given by \cite{TakahashiBoyd1987, Takahashi1983, Takahashi1987},

\begin{equation}
\lambda = \frac{\ln 2}{f t_{1/2}}\, f^{*}.
\label{eq1}
\end{equation}

Here, $t_{1/2}$ is the partial half-life of the specific parent-daughter energy level combination for which a transition rate has to be calculated and $f_a^{*}$ is the lepton phase volume. For allowed  $\beta^{-}$ decay, the expression for the decay rate function $f(Z,W_0)$ can be simplified to \cite{GoveMartin1971, konopinski1941}

\begin{equation}
f(Z,W_0) = \int_{1}^{W_0} \sqrt{W^2-1}\,W\,(W_0-W)^2 F_0(Z,W)L_0\,dW .
\label{eq2}
\end{equation}

The certain combinations of electron radial wave functions evaluated at the nuclear radius R is given by the $L_0$. The term $F_0 (Z, W )$ is the Fermi function. More details of the lepton phase volume can be found in Ref. \cite{Gupta2019}. The lepton phase volume $f^{*}$ \cite{Takahashi1983} for the continuum state
$\beta^{-}$ decay can be expressed as

\begin{equation}
f^{*}(\mathrm{continuum}) = \int_{1}^{W_c} \sqrt{W^{2}-1}\,W\,(W_c-W)^{2}
F_0(Z,W)L_0\,dW,
\label{eq3}
\end{equation}

with $W_c = Q_c/m_e c^2 + 1$ is the maximum energy available to the emitted $\beta^{-}$ particle. The continuum decay Q-value $Q_c$ is given by

\begin{equation}
Q_c = Q_n - \left[B_n(Z+1)-B_n(Z)\right].
\label{eq4}
\end{equation}

The term $\left[B_n(Z+1)-B_n(Z)\right]$ denotes the difference of binding energies for bound electrons of the daughter and the parent atom. The experimental values for all the atomic data (binding energies or ionisation potentials) are available in Ref. \cite{NIST}. Furthermore, for bound state $\beta^{-}$ decay of the bare atom, $f^{*}$ takes the form \cite{Takahashi1983}

\begin{equation}
f^{*}(\mathrm{bound}) = \sum_{x}
\sigma_x \left(\frac{\pi}{2}\right)
\left[f_x \ \mathrm{or}\ g_x\right]^2 b^2,
\qquad
(x = ns_{1/2}).
\label{eq5}
\end{equation}

Here, $[f_x \text{ or } g_x]$ is the larger component of the electron radial wave function evaluated at the nuclear radius $R$ of the daughter for the orbit $x$. $[f_x \text{ or } g_x]$ is obtained by solving the Dirac radial wave equations. In our case, $\sigma_x$ is the vacancy of the orbit, chosen as unity, and $b$ is equal to $Q_b/m_e c^2$ where

\begin{equation}
Q_b = Q_n - \left[B_n(Z+1)-B_n(Z)\right] + B_{\mathrm{shell}}(Z+1).
\label{eq6}
\end{equation}

Thus, in case of bare atom the total decay rate becomes
\vspace{-0.2cm}
\begin{align}
\lambda_{\mathrm{bare}}
 &= \lambda_{\mathrm{bound}} + \lambda_{\mathrm{continuum}}, \\
\text{i.e.,}\qquad
\lambda_{\mathrm{bare}}
 &= \frac{\ln 2}{f t_{1/2}}
 \left[ f^{*}(\mathrm{bound}) + f^{*}(\mathrm{continuum})
 \right].
\end{align}
    
Finally, the bare to neutral decay rate can be obtained from,

\begin{equation}
    \frac{\lambda_{bare}}{\lambda_{neutral}} = \frac{f^{*}(\mathrm{bound}) + f^{*}(\mathrm{continuum})}{f}.
\end{equation}

\subsection{\label{machinelearning} Machine Learning Framework}

In this work, two complementary approaches are employed to incorporate machine learning techniques for estimating the ratio $\lambda_{\mathrm{bare}}/\lambda_{\mathrm{continuum}}$. In the first approach, symbolic regression (SR) and Gaussian process regression (GPR) are used to derive a data driven parametrisation of the decay rate ratio. In the second approach, artificial intelligence and machine learning (AIML) models, namely neural networks (NN) and random forests (RF), are developed to predict the bare to neutral $\beta^{-}$ decay rate ratio directly from the underlying nuclear properties.

\subsubsection{\label{parametrisation} Data Driven Parametrisation}

For each $\beta^-$ active radionuclide, several transitions may occur from a given parent level to different daughter levels. Each of these transitions is characterised by a unique Q-value, $Q_n$. In case of bare atom $\beta^-$ decay to continuum as well as to the atomic bound state may occur, as discussed earlier. Both the bound and continuum decay rates depend strongly on $Q_n$, along with other nuclear parameters such as the mass number A and the atomic number Z. Consequently, for transitions originating from a fixed parent nucleus, the ratio of the total decay rate in the case of the bare atom (i.e. the sum of the bound and continuum decay rate) to  that of the neutral atom (terrestrial decay rate) is not constant; rather, it increases as the transition $Q_n$ decreases \cite{Gupta2019}.  

This observable was treated as the dependent variable and analysed as a function of the neutral atom $Q_n$, for fixed values of the mass number A and atomic number Z. However, for each A, Z combination, only a limited number of discrete data are available. To overcome this issue, we have performed a two-step regression strategy based on artificial intelligence and machine learning (AIML). 

Gaussian Process Regression was used as the first step of the analysis to reconstruct a smooth underlying dependence on $Q_n$ from the sparse data. GPR is a non-parametric regression technique that models the data as a realization of a Gaussian process defined by a covariance kernel, which encodes smoothness of the curve \cite{Rasmussen2003}. GP defines a probability distribution over possible functions rather than fitting deterministic curve, and mathematically represented as

\begin{equation}
f(Q_n) \sim GP\big(m(Q_n),\, k(Q_n, Q_n')\big),
\end{equation}

where the mean function $m(Q_n)$ is assumed to be zero, representing a neutral prior. The kernel function $k(Q_n, Q_n')$ encodes the physical smoothness of the observable. It captures correlations between nearby $Q_n$ values and provides uncertainty estimates for the predictions. A radial basis function (RBF) kernel, multiplied by a constant kernel, is employed to enforce smooth variation with $Q_n$.

\begin{equation}
k(Q_n, Q_n') = \sigma_f^2 \exp\left(-\frac{(Q_n - Q_n')^2}{2 l^2}\right).
\end{equation}

The kernel hyperparameters were optimised via marginal likelihood maximisation. Once trained on the original data points, the Gaussian Process Regression (GPR) model was evaluated on a dense grid of $Q_n$ values within the data range, generating a smooth set of interpolated points. This procedure enabled the creation of additional data points for subsequent fitting by capturing the global trend of the observable.

Symbolic regression was then employed to extract an explicit mathematical expression describing the $Q_n$ dependence of each observable~\cite{Schmidt2009}. Unlike conventional regression methods that rely on predefined functional forms, symbolic regression searches directly over mathematical expressions by constructing candidate functions

\begin{equation}
y(Q_n) \approx F(Q_n; \theta).
\end{equation}

Where, the expression $F$ is constructed from a predefined operator set motivated by physical considerations. However, in low-data regimes, symbolic regression (SR) is prone to overfitting~\cite{Cranmer2020}. To mitigate this issue, SR was applied to a combined dataset consisting of the original data points and the GPR-generated points, with higher weights assigned to the original measurements. This ensured that the resulting expressions remained anchored to the true data, while the GPR output served only to constrain the overall functional form. Furthermore, the expression complexity was restricted to obtain physically meaningful representations. This procedure yielded stable analytical expressions for the observable across different nuclei and their isotopes.

\subsubsection{\label{machinelearning} Machine Learning Prediction}

We have used two supervised regression algorithms — Random Forest (RF) and Artificial Neural Network (ANN) to predict the $\lambda_{bare}/ \lambda_{neutral}$ using three nuclear input features: the Q-value ($Q_n$), mass number ($A$), 
and daughter atomic number ($Z$).

\paragraph{Data Preparation}
The machine learning problem is formally defined over a dataset $D = {(x_i, y_i)}$, where each input vector $x_i = (Q_n, A, Z)$ represents the nuclear features, and the target variable $y_i$ represents the decay rate ratio $\lambda_{bare} / \lambda_{neutral}$. The dataset was partitioned into training (80\%) and test (20\%) subsets using stratified sampling based on binned $Z$ and $Q_n$ quartiles, ensuring representative coverage of the feature space \cite{pedregosa2011}. Outliers in the training set were identified and removed via the interquartile range (IQR) criterion a sample was excluded if any feature $x_j$ satisfied \cite{tukey1977}
\begin{equation}
    x_j < Q_1^{(j)} - 1.5\,\text{IQR}^{(j)} 
    \quad \text{or} \quad 
    x_j > Q_3^{(j)} + 1.5\,\text{IQR}^{(j)},
\end{equation}
where $Q_1^{(j)}$ and $Q_3^{(j)}$ denote the first and third quartiles of feature $j$, respectively. The features were subsequently normalised using a robust scaler centred on the median \cite{pedregosa2011}. Where, robust scaling is given by

\begin{equation}
    x' = \frac{x-Median}{IQR},
\end{equation}

with $x'$ is the scaled data point, and x is the original data point. Because the target spans several orders of magnitude, a logarithmic transformation \cite{box1964} $\tilde{y} = \ln(1 + y)$ was applied before training; predictions were mapped back via $y = e^{\hat{\tilde{y}}} - 1$.

\paragraph{Models and Hyperparameters}

To map the non-linear relationships within the data, we deploy a Multi-Layer Perceptron neural network \cite{rumelhart1986, hornik1989}. This model functions as a highly flexible function approximator. The architecture consists of multiple hidden layers that progressively transform the input features. Each hidden layer l computes its output via the operation

\begin{equation}
h^{(l)} = \sigma \!\left( W'^{(l)} h^{(l-1)} + b^{(l)} \right)
\end{equation}

In this equation, W' represents the learned weight matrix and b is the bias vector. We utilize the Rectified Linear Unit \cite{nair2010} for the activation function $\sigma$. The ReLU function is defined mathematically as  $\sigma$(z) = max(0, z). This activation introduces necessary non-linearity without causing the gradients to vanish during training. Optimization is driven by the Adam algorithm \cite{kingma2015}. Adam adaptively scales the learning rate based on historical gradient moments to efficiently minimize the Mean Squared Error. The MLP-NN consisted of two fully connected hidden 
layers of 128 and 64 neurons with ReLU activations,
\begin{equation}
    f(\mathbf{x}) =
    \max\left(0,\, \mathbf{W}'\mathbf{x} + \mathbf{b}\right).
\end{equation}
trained with the Adam optimiser (up to 3000 epochs) and early stopping monitored 
on a held-out validation subset.

We have performed a Random Forest (RF) \cite{breiman2001} regressor in parallel to provide a complementary, tree based modeling approach. This algorithm constructs an ensemble of independent decision trees. Each individual tree is trained on a randomized bootstrap sample of the dataset. During training, the trees partition the feature space using threshold-based rules. The final prediction of the forest is derived by averaging the outputs of all the individual trees. This ensemble averaging process fundamentally reduces the overall variance of the model. It prevents the system from overfitting to the statistical noise inherent in limited nuclear datasets. The forest is particularly valuable here because it does not require strict feature scaling. It also provides a direct mechanism to assess feature importance across the different decay regimes. The RF regressor consisted of $N_\text{tree} = 200$ decision trees with maximum depth 
$d_\text{max} = 12$, minimum leaf size $n_\text{leaf} = 5$, and $\sqrt{p}$ features 
considered at each split ($p = 3$). Finally, to ensure the strict reproducibility of all data splitting, bootstrap sampling, and weight initialization procedures, a constant random seed was fixed across all implemented algorithms.

\begin{figure}[p]
\centering
\resizebox{\textwidth}{!}{%
\begin{tikzpicture}[
font=\normalsize,
every node/.style={align=center, text=black},
arr/.style={-{Triangle[length=3mm,width=2mm]}, flowcol, thick},
line/.style={flowcol, thick},
node distance=6mm
]

\node[text width=10cm] (top) {\bfseries Theoretical $\beta^{-}$ decay calculations\\
\normalsize (Phase space for bare \& neutral atoms; input: $Q_n,\,A,\,Z$)};

\node[text width=8cm,below=of top] (set1) {\bfseries SET-1 (Training Dataset)\\
\normalfont Calculated hypothetical $\beta^-$ transitions\\
Target: $\lambda_{\rm bare}/\lambda_{\rm neutral}$};
\draw[arr] (top.south) -- (set1.north);

\coordinate (splitTop) at ($(set1.south)+(0,-6mm)$);
\coordinate (leftX) at (-4.6,0);
\coordinate (rightX) at (4.6,0);
\draw[line] (set1.south) -- (splitTop);
\draw[line] (splitTop -| leftX) -- (splitTop -| rightX);
\draw[arr] (splitTop -| leftX) -- ++(0,-6mm);
\draw[arr] (splitTop -| rightX) -- ++(0,-6mm);

\node[text width=4.2cm,font=\bfseries, below=8mm of splitTop, xshift=-4.6cm, anchor=north] (apI)
  {Approach I:\\ Data driven Parametrisation};
\node[text width=4.2cm,font=\bfseries, below=8mm of splitTop, xshift=4.6cm, anchor=north] (apII)
  {Approach II:\\ Machine Learning Prediction};

\node[text width=4.2cm,below=of apI] (gpr) {Gaussian Process Regression\\ (RBF kernel)};
\draw[arr] (apI.south) -- (gpr.north);

\node[text width=4.2cm,below=of gpr] (dense) {Generate dense $Q_n$ interpolation};
\draw[arr] (gpr.south) -- (dense.north);

\node[text width=4.2cm,below=of dense] (merge) {Weighted dataset\\ (Original data + GPR interpolation)};
\draw[arr] (dense.south) -- (merge.north);

\node[text width=4.2cm,below=of merge] (sr) {Symbolic Regression};
\draw[arr] (merge.south) -- (sr.north);

\node[text width=4.2cm,below=of sr] (coeff) {Generalise coefficients\\
fixed $(A,Z)$ $\rightarrow$ fixed $Z$ $\rightarrow$ global $(A,Z)$};
\draw[arr] (sr.south) -- (coeff.north);

\node[text width=4.4cm,below=of coeff] (formula) {Final analytical equation\\ $\mathbf{F(Q_n,A,Z)}$};
\draw[arr] (coeff.south) -- (formula.north);

\node[text width=4.2cm,below=of apII] (prep) {Shared preprocessing\\
(80/20 split, IQR outlier removal, robust scaling, log transform)};
\draw[arr] (apII.south) -- (prep.north);

\coordinate (splitPrep) at ($(prep.south)+(0,-6mm)$);
\coordinate (nnX) at (2.9,0);
\coordinate (rfX) at (6.3,0);
\draw[line] (prep.south) -- (splitPrep);
\draw[line] (splitPrep -| nnX) -- (splitPrep -| rfX);
\draw[arr] (splitPrep -| nnX) -- ++(0,-6mm);
\draw[arr] (splitPrep -| rfX) -- ++(0,-6mm);

\node[text width=3.2cm, below=8mm of splitPrep, xshift=-1.7cm, anchor=north] (nn)
  {Neural Network\\ MLP (128--64)\\ Adam optimiser};
\node[text width=3.2cm, below=8mm of splitPrep, xshift=1.7cm, anchor=north] (rf)
  {Random Forest\\ 200 trees\\ Depth = 12};

\node[text width=3.2cm,below=of nn] (nnpred) {NN prediction\\ $f(Q_n,A,Z)$};
\node[text width=3.2cm,below=of rf] (rfpred) {RF prediction\\ $f(Q_n,A,Z)$};
\draw[arr] (nn.south) -- (nnpred.north);
\draw[arr] (rf.south) -- (rfpred.north);

\coordinate (mlBar) at ($(nnpred.south)!0.5!(rfpred.south)+(0,-6mm)$);
\draw[line] (nnpred.south) -- (nnpred.south |- mlBar);
\draw[line] (rfpred.south) -- (rfpred.south |- mlBar);
\draw[line] (nnpred.south |- mlBar) -- (rfpred.south |- mlBar);
\coordinate (mlEnd) at ($(mlBar)+(0,-20mm)$);
\draw[arr] (mlBar) -- (mlEnd);

\coordinate (mergeBar) at ($(formula.south)!0.5!(mlEnd)+(0,-10mm)$);
\draw[line] (formula.south) -- (formula.south |- mergeBar);
\draw[line] (mlEnd) -- (mlEnd |- mergeBar);
\draw[line] (formula.south |- mergeBar) -- (mlEnd |- mergeBar);
\draw[arr] (mergeBar) -- ++(0,-6mm);

\node[text width=8cm, below=8mm of mergeBar, anchor=north] (valid) {\bfseries Validation (SET-2)\\
\normalfont Using calculated $\lambda_{\rm bare}/\lambda_{\rm neutral}$ on real transitions\\
Performance metrics: $R^2$, MAE};

\node[text width=8.5cm,below=8mm of valid] (set3) {\bfseries Application (SET-3)\\
\normalfont Prediction of $\lambda_{\rm bare}/\lambda_{\rm neutral}$ for astrophysical nuclei};
\draw[arr] (valid.south) -- (set3.north);

\end{tikzpicture}
}
\caption{Overall workflow adopted in the present work.}
\label{fig:workflow}
\end{figure}
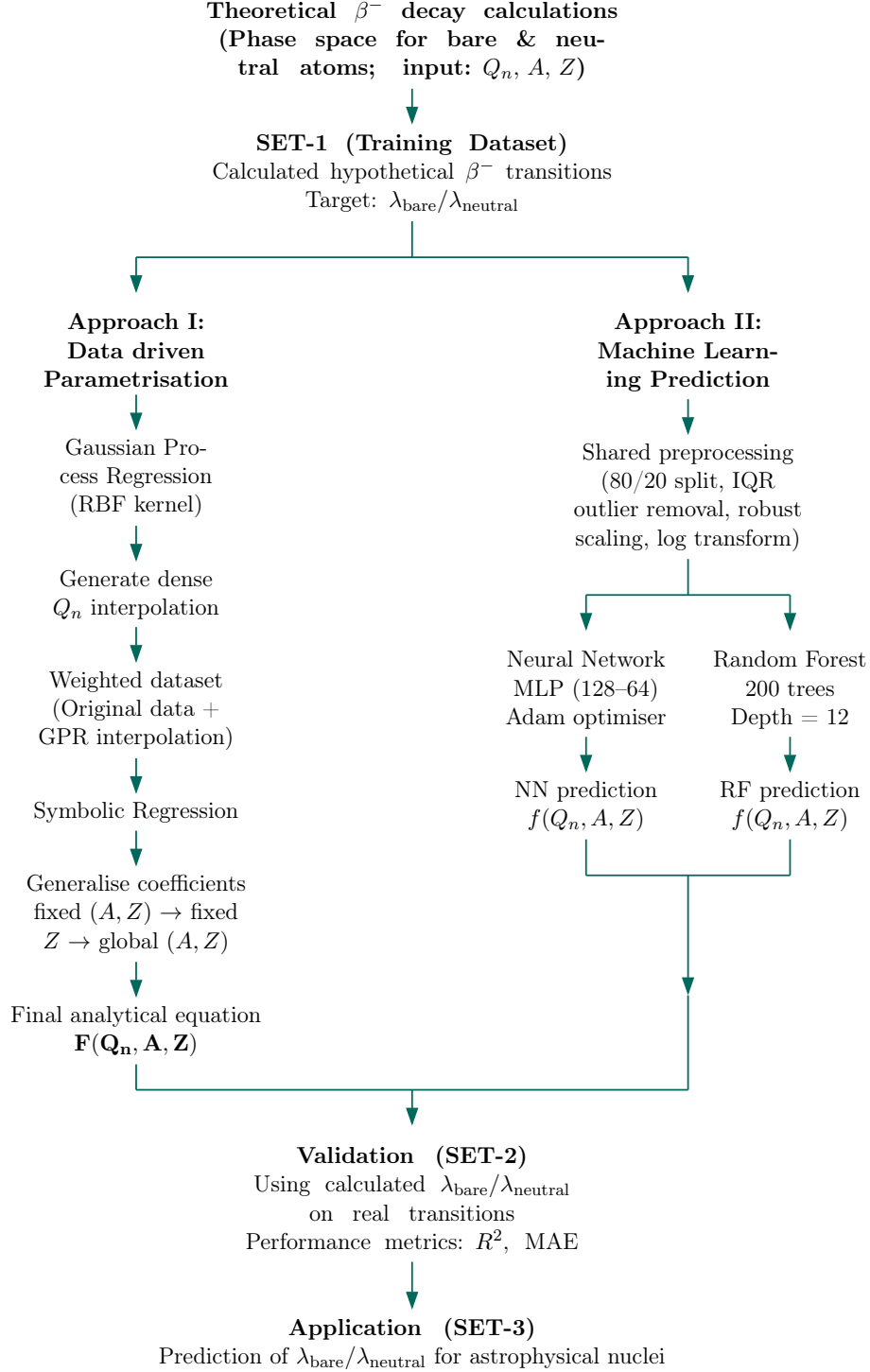

\paragraph{Performance Metrics}
Model performance was quantified by the coefficient of determination $R^2$, and mean absolute error (MAE). The coefficient of determination, \(R^2\), quantifies the fraction of the variance in the observed data that is explained by the model:

\begin{equation}
R^2
=
1
-
\frac{\sum_{i=1}^{N}(y_i-\hat{y}_i)^2}
{\sum_{i=1}^{N}(y_i-\bar{y})^2},
\label{eq:r2}
\end{equation}

where \(\bar{y}\) is the mean of the observed values,  \(y_i\) denotes the observed value, and \(\hat{y}_i\) represents the corresponding predicted value. An \(R^2\) value close to unity indicates excellent agreement between the model predictions and the observed data. Whereas, the mean absolute error (MAE) measures the average magnitude of the prediction errors and is defined as

\begin{equation}
\mathrm{MAE}
=
\frac{1}{N}
\sum_{i=1}^{N}
\left| y_i - \hat{y}_i \right|,
\label{eq:mae}
\end{equation}

where \(N\) is the total number of data points. Smaller values of MAE indicate better predictive accuracy, with \(\mathrm{MAE}=0\) corresponding to perfect agreement between predictions and observations. the overall methodology adopted in this work is presented with the flowchart in Fig. \ref{fig:workflow}.

\section{\label{data sources} Data Sources}
Producing fully ionised atoms in a terrestrial laboratory is experimentally challenging. Consequently, only a limited number of experiments have been performed to date~\cite{Jung1992, Bosch1996}, and among these, bound and continuum state $\beta^-$ decay rates have been measured separately for only one nucleus, $^{207}$Tl~\cite{Ohtsubo2005}. Therefore, reliable estimates of $\lambda_b$ and $\lambda_c$ must predominantly rely on theoretical calculations. In the present work, we employ the formalism developed in Ref.~\cite{Gupta2019}, where the bound state and continuum state decay rates, $\lambda_b$ and $\lambda_c$, were calculated for fully ionised atoms under terrestrial conditions. The corresponding phase space volumes for bare and neutral atoms were evaluated for allowed $\beta^-$ transitions, enabling a direct comparison between the two cases. Since the absence of orbital electrons maximises the contribution from bound state decay channels, the resulting calculations provide the upper limit of $\lambda_{\mathrm{bare}}$ and, consequently, of the ratio $\lambda_{\mathrm{bare}}/\lambda_{\mathrm{neutral}}$.


A fundamental requirement for training machine learning models is the availability of data that adequately sample the full range of input parameters. However, in practice, only a limited number of $\beta^-$ transitions are available for any given nucleus, leading to uneven coverage of the $(Z, A, Q_n)$ parameter space. To address this limitation, we generated a large set of hypothetical transitions (2271 in number) and calculated the corresponding phase space volumes for both bare and neutral atoms. Hereafter, this dataset is designated as Set-1. It should be emphasised that this approach does not compromise the underlying physics: the $\beta^-$ decay rate depends on both the phase space volume and the overlap of the parent and daughter nuclear wavefunctions, but the phase space volume itself is independent of the latter. Since our primary quantity of interest is the ratio $\lambda_{\mathrm{bare}}/\lambda_{\mathrm{neutral}}$ for each transition---a ratio that is entirely governed by the phase space volume---the inclusion of hypothetical transitions is physically well justified.

To assess the predictive capability of the obtained parametrisations and the AIML models, we constructed a second dataset (consists of 151 transitions), denoted as Set-2, consisting of experimentally known $\beta^-$ transitions from selected radioactive nuclei. For these nuclei, previously calculated bare atom and neutral atom decay rates are available. The corresponding values of $\lambda_{\mathrm{bare}}/\lambda_{\mathrm{neutral}}$ obtained from the fitted equations and AIML models are compared with the calculated results to validate the proposed methodology.

Finally, the fitted parametric equation and AIML models were applied to a set of nuclei of astrophysical interest to predict the ratio $\lambda_{\mathrm{bare}}/\lambda_{\mathrm{neutral}}$. This dataset is hereafter referred to as Set-3, comprises nearly 600 allowed transitions in the mass range $A = 59 - 210$ and includes nuclei of astrophysical interest that are potential candidates for future storage ring experiments. The resulting predictions provide estimates of the maximum possible enhancement of the $\beta^-$ decay rate arising from complete ionisation and may serve as useful inputs for astrophysical applications involving highly ionised stellar environments.



\section{\label{results}Results and Discussion}
\subsection{\textbf{Data-Driven Parametrisation}}

In this section, we discuss our investigation on the systematic behaviour of $\beta^-$ decay ratio of bare to neutral atoms, $(\lambda_{bare} / \lambda_{neutral})$, using machine learning (ML) assisted regression techniques. A combination of Gaussian Process Regression (GPR) and Symbolic Regression (SR) is employed to identify an interpretable functional dependence on nuclear parameters as discussed earlier.


\begin{figure}[H]
\centering

\includegraphics[width=0.85\textwidth]{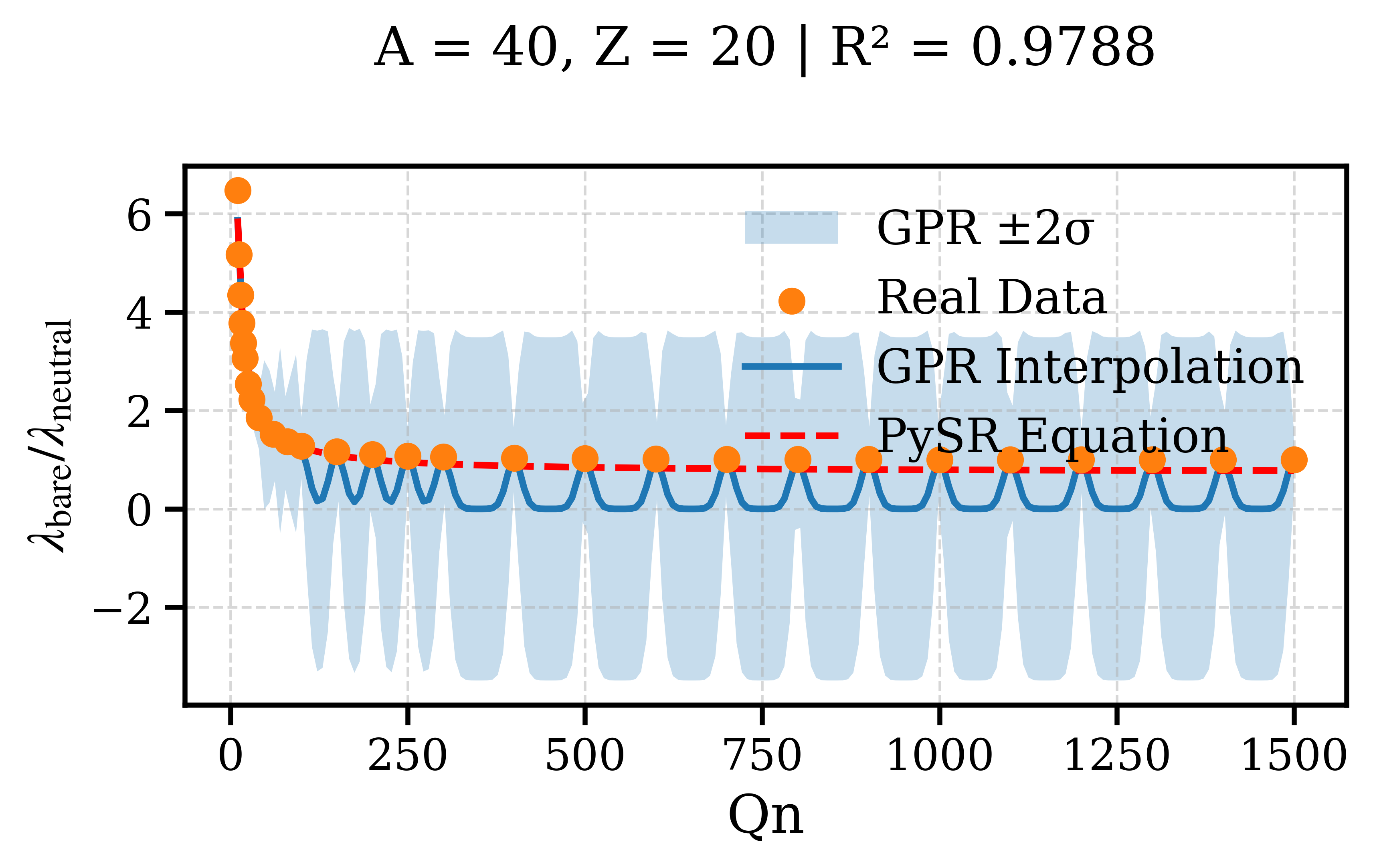}

\vspace{0.1cm}

\includegraphics[width=0.85\textwidth]{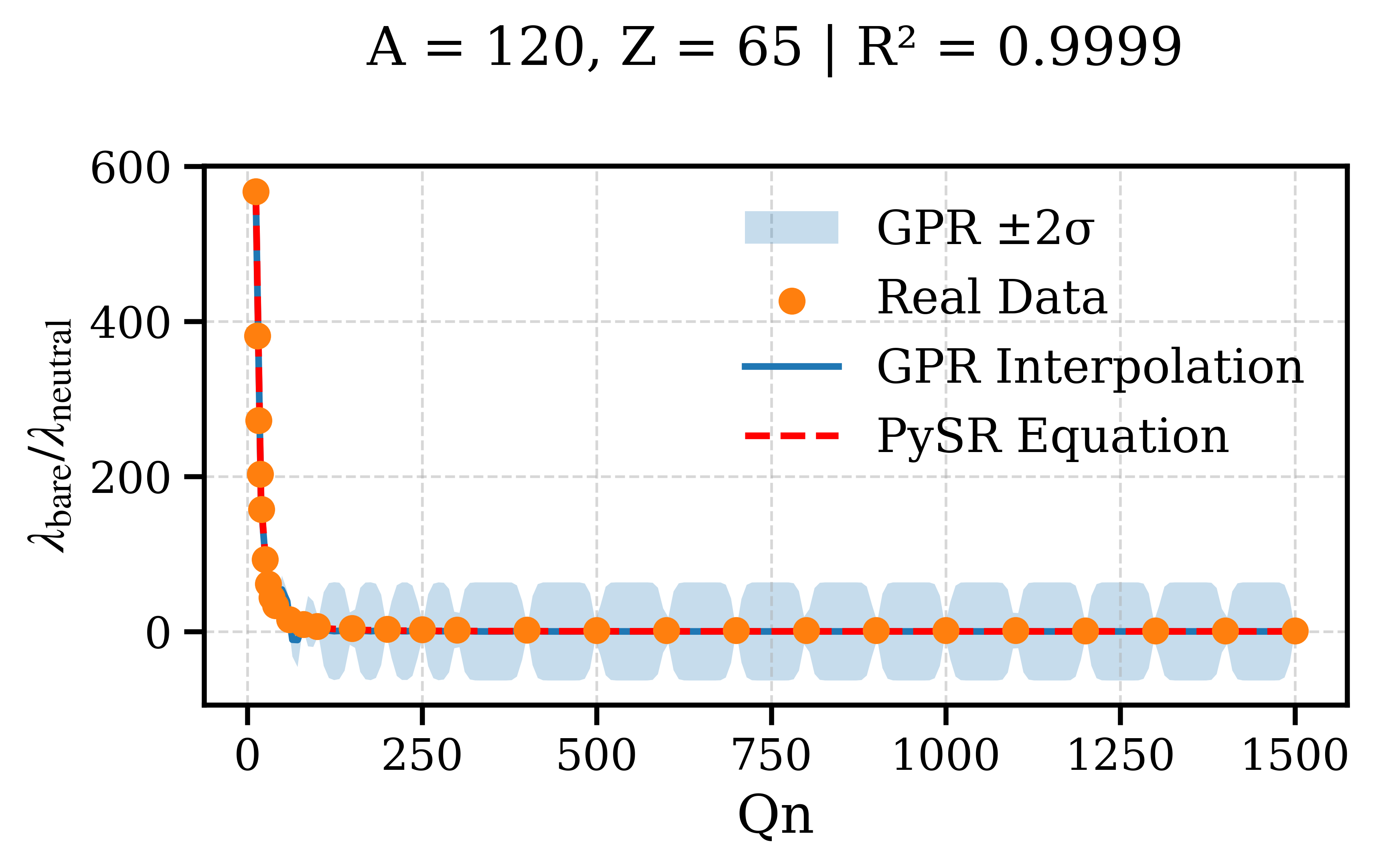}

\caption{
Comparison of GPR interpolation and SR fits for representative nuclei:
(a) $A=40$, $Z=20$ and
(b) $A=120$, $Z=65$.
The orange points denote the tabulated data, the blue solid line represents the GPR interpolation, and the red dashed line corresponds to the symbolic regression (SR) expression. The shaded region indicates the $2\sigma$ uncertainty band of the GPR prediction.
}
\label{fig:GPRSR}
\end{figure}

\begin{figure}[p]
\ContinuedFloat
\centering

\includegraphics[width=0.85\textwidth]{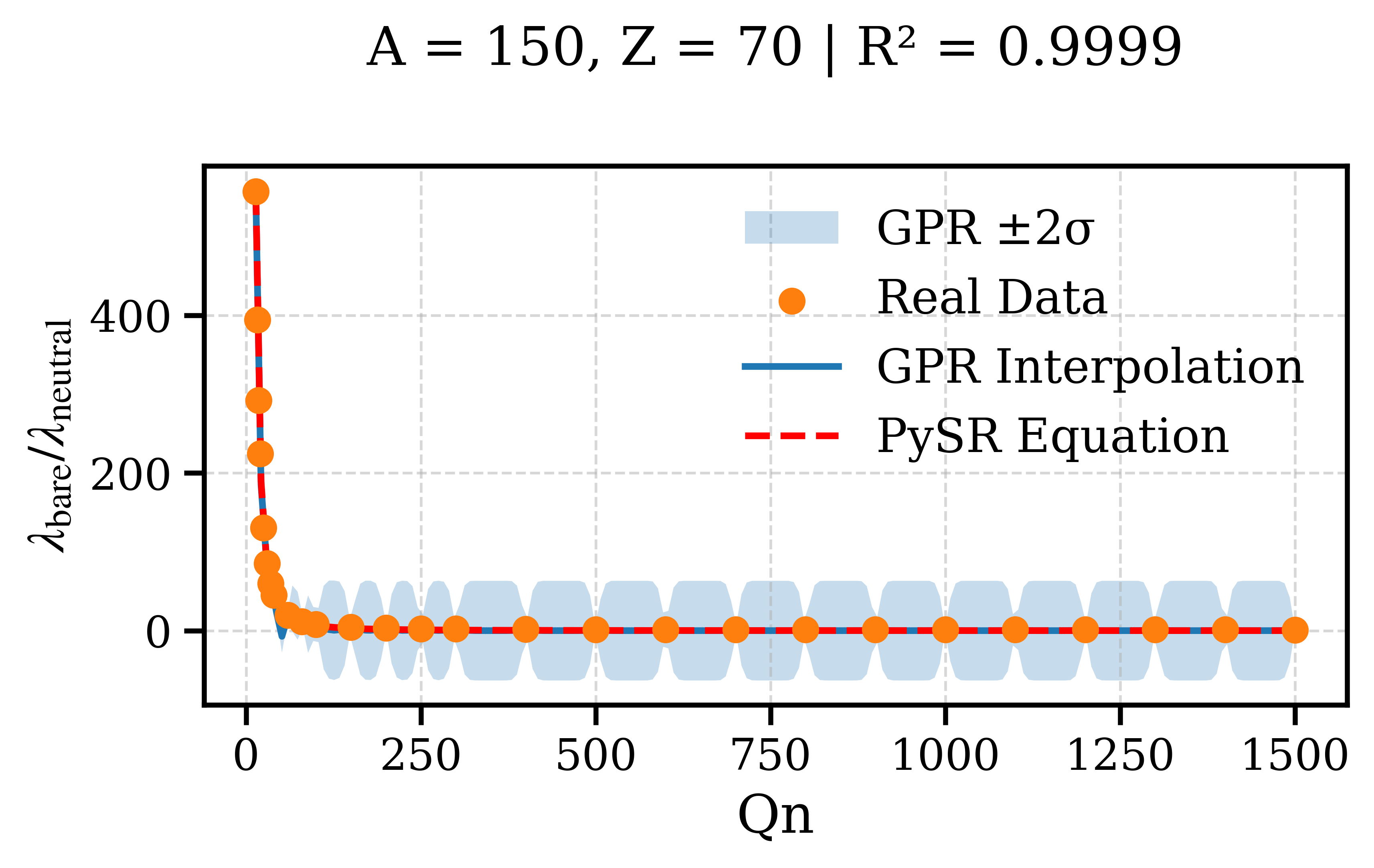}

\vspace{0.1cm}

\includegraphics[width=0.85\textwidth]{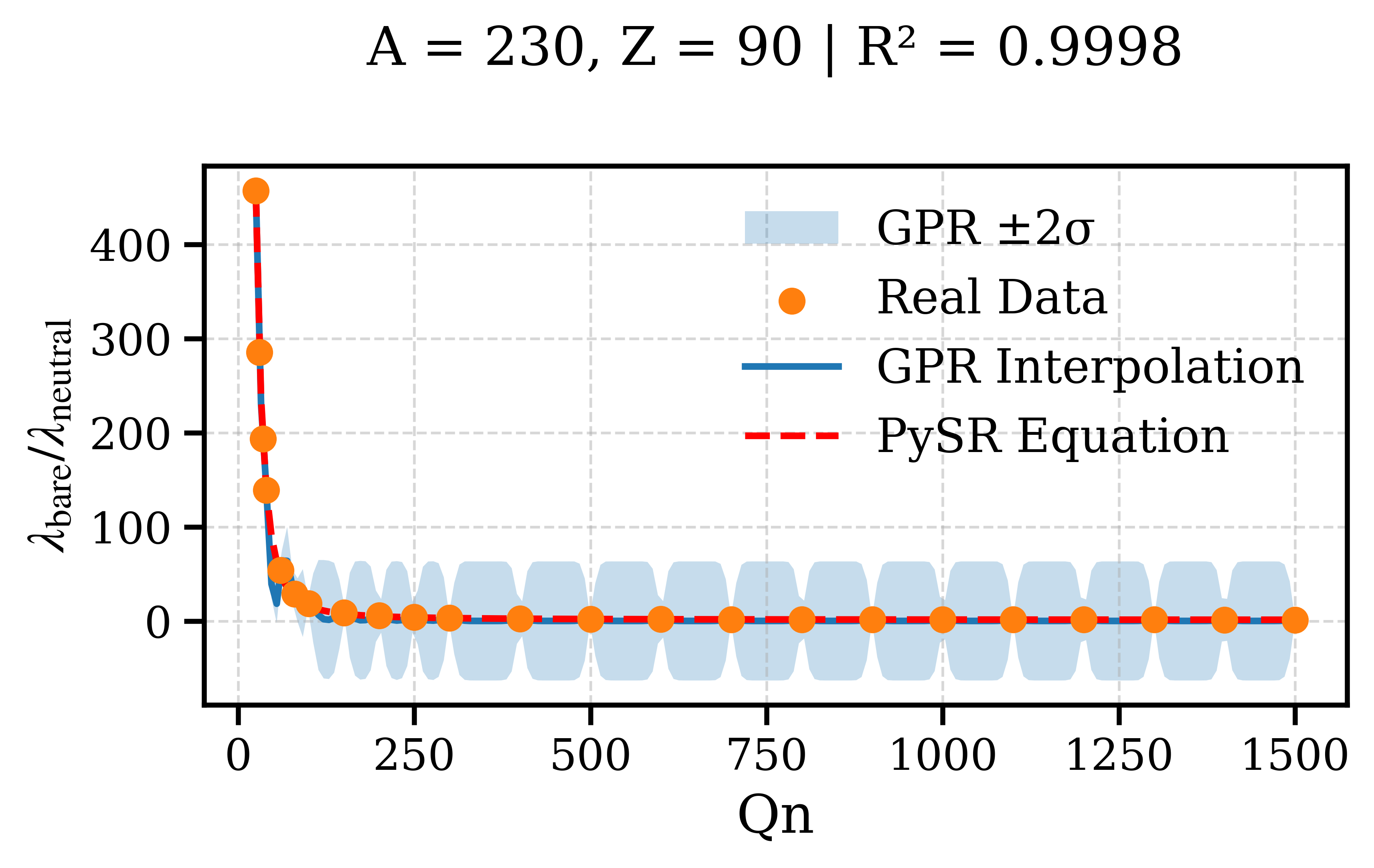}

\caption[]{Continued.
Comparison of GPR interpolation and SR fits for representative nuclei (continued):
(c) $A=150$, $Z=70$ and
(d) $A=230$, $Z=90$.
}
\end{figure}

\subsubsection{\textbf{Functional dependence on $Q_n$}}

\begin{figure}[htbp]
\centering
\includegraphics[width=1.0\textwidth]{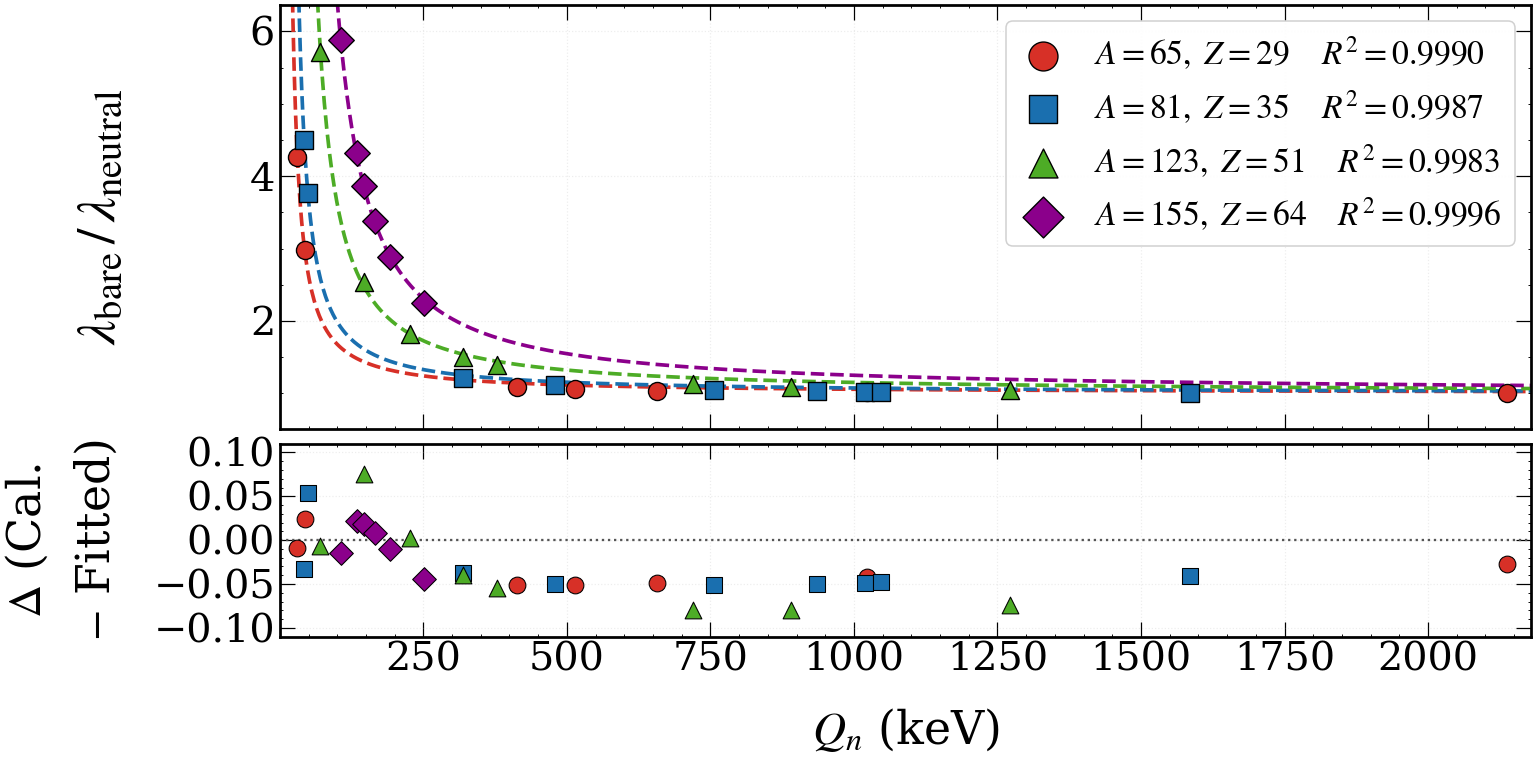}
\caption{Validation of the SR parametrisation using nuclei from Data Set-2. In the upper panel, the fitted curves accurately reproduce the calculated $\lambda_{\mathrm{bare}}/\lambda_{\mathrm{neutral}}$ ratios over a wide range of decay energies, for different nuclei, demonstrating the predictive capability and robustness of the proposed functional form. The lower panel shows the deviation of the fitted values from the corresponding theoretical values.}
\label{set2_fitting}
\end{figure}

To systematically analyse the behaviour of this ratio, we first employ GPR method, on each of the nuclei of Set-1, in order to generate a dense and smooth representation of the available discrete data. The GPR framework not only interpolates between the calculated data points but also provides an estimate of the associated uncertainty, thereby allowing us to assess the reliability of the interpolation. From this procedure, a mean interpolated dataset is obtained, as illustrated in Fig. \ref{fig:GPRSR}. These interpolated data points, shown with blue curve, along with the $\pm 2\sigma$ uncertainty - shown with blue shaded region. All these points are then used as input for SR, which aims to identify an explicit analytical expression describing the underlying trend. The SR predicted trend is also shown in Fig. \ref{fig:GPRSR} and reveals that the $(\lambda_{bare} / \lambda_{neutral})$ follows an exponential law dependence on the decay energy $Q_n$ as 


\begin{equation}
\frac{\lambda_{bare}}{\lambda_{neutral}} = \exp\left(\frac{c}{x + x_0}\right),
\end{equation}

where $x \equiv Q_n$, and $c$ and $x_0$ are fitting parameters. This functional form captures the strong non-linear sensitivity of the decay channels to the available phase space. The SR equation, shown by the dashed line, reproduces the calculated data with excellent accuracy, yielding coefficients of determination ($R^2$) in the range of 0.97–0.99 across different nuclei and mass regions, as illustrated in Fig.~\ref{fig:GPRSR}. To further validate the proposed functional form, the same parametrisation was applied to Data Set-2 (Fig. \ref{set2_fitting}). The resulting fits also exhibit excellent agreement, with $R^2$ values approaching unity, thereby confirming the robustness and broad applicability of the SR based parametrisation. At this stage, the parameter space is characterised by the four quantities $(A, Z, c, x_0)$.

\paragraph{Determination of $c(Z, A)$ and $x_0(Z,A)$}

To investigate the global systematics of the parameters $c$ and $x_0$, we first fix the proton number $Z$ and study their variation with the mass number $A$ as shown in Fig. \ref{set2_fitting}. For a given isotopic chain, both parameters exhibit an approximately linear dependence on $A$, which can be expressed as

\begin{eqnarray}
c(A) &= c_{\mathrm{intercept}} + A \cdot c_{\mathrm{slope}}, \
x_0(A) &= x_{0,\mathrm{intercept}} + A \cdot x_{0,\mathrm{slope}}.
\end{eqnarray}

\begin{figure}[H]
\centering

\includegraphics[width=0.85\textwidth]{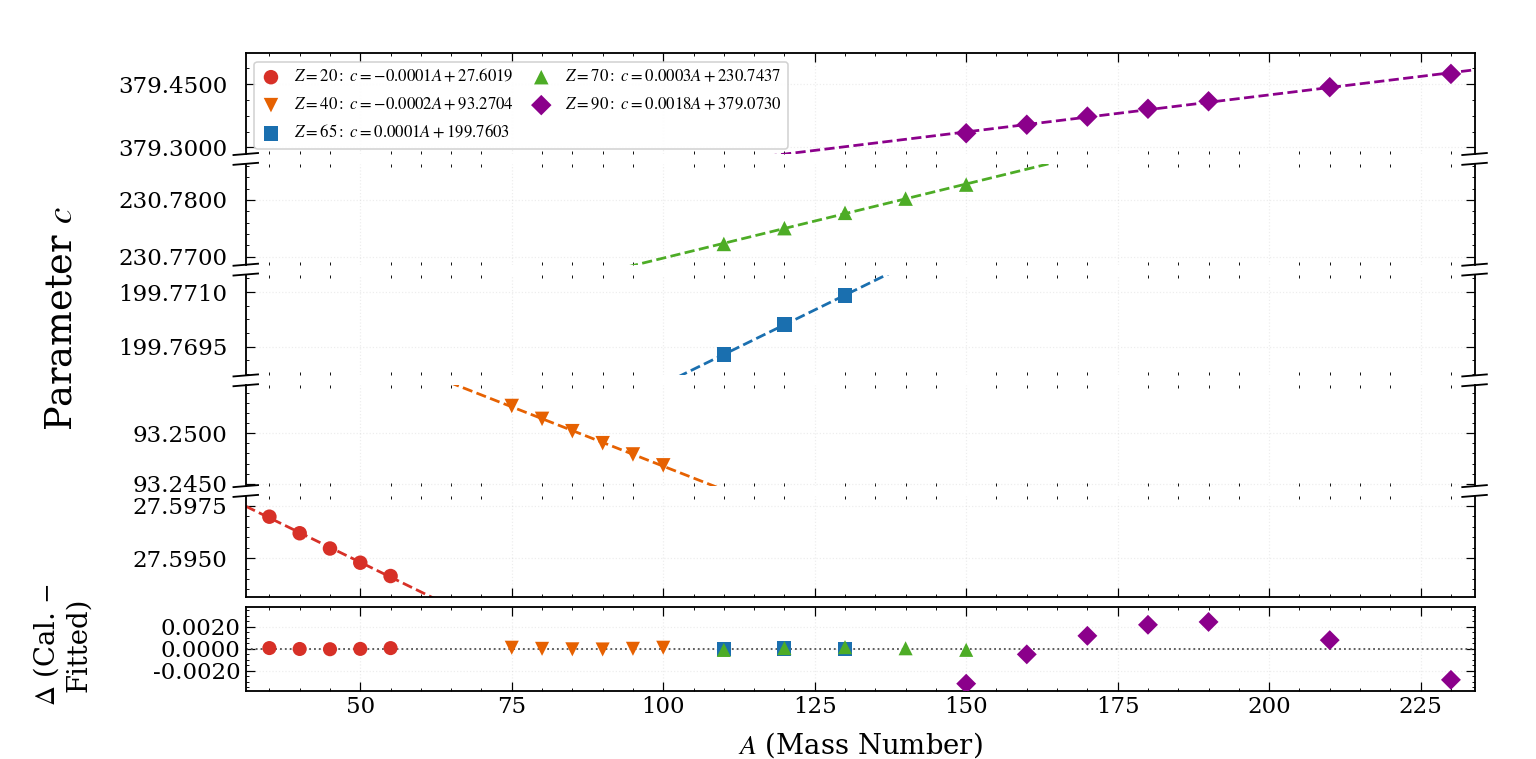}

\vspace{0.1cm}

\includegraphics[width=0.85\textwidth]{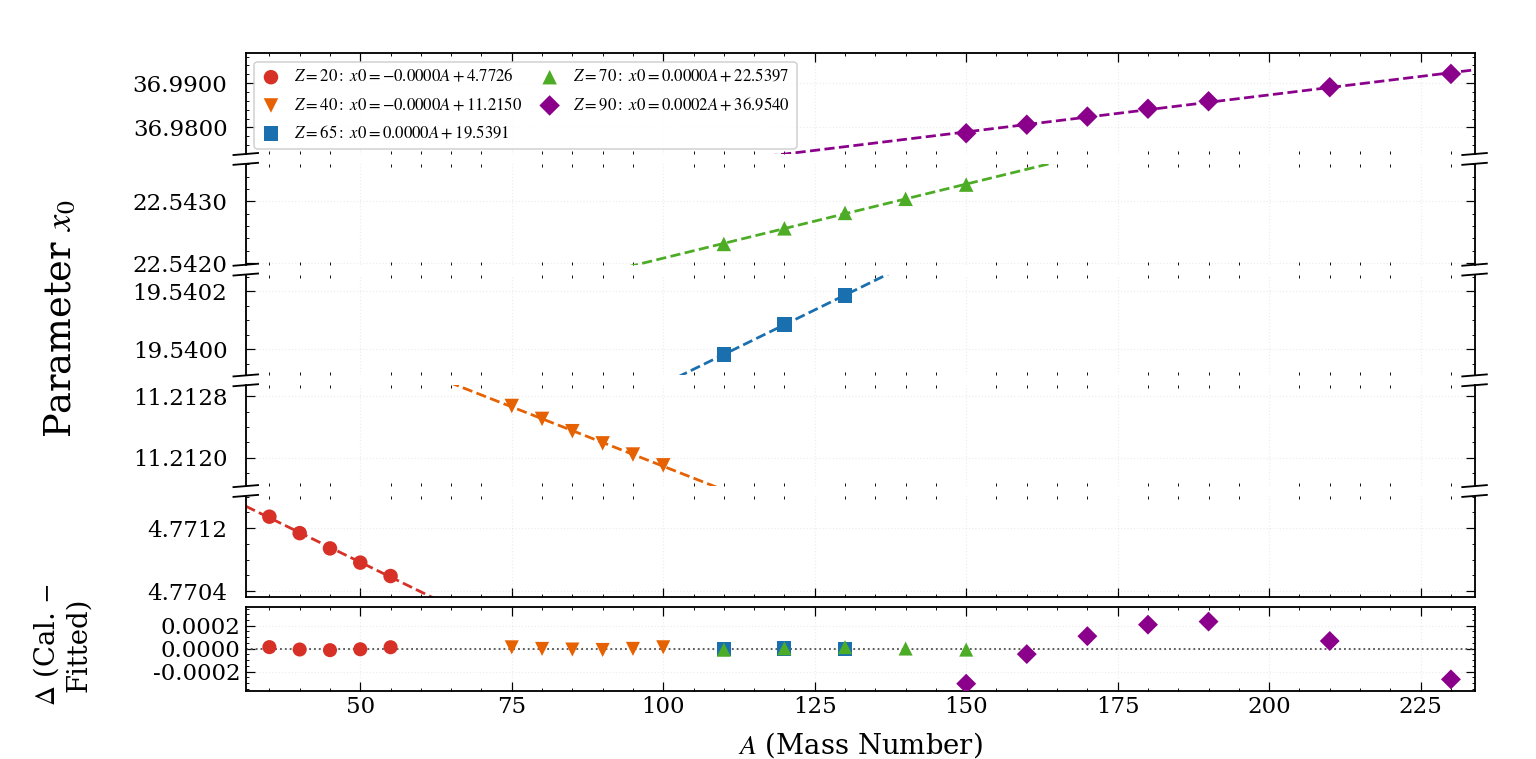}

\caption{Linear fits of the SR parameters (top) $c$ and (bottom) $x_0$ as functions of mass number $A$ for different isotopic chains. The extracted slopes and intercepts are subsequently used to investigate the systematic dependence of the fitting parameters on proton number $Z$. The lower panel of each plot shows the deviation between the fitted and original values.}
\label{fig:c_intercept_slope}
\end{figure}

Representative linear fits are shown in Fig.~\ref{fig:c_intercept_slope}. The extracted slope parameters, $c_{\mathrm{slope}}$ and $x_{0,\mathrm{slope}}$, are generally small in magnitude, indicating that the dependence of $c$ and $x_0$ on the mass number is relatively weak. Nevertheless, a systematic evolution is observed with increasing proton number. In particular, both slope parameters gradually change sign from negative to positive values as $Z$ increases (Z $\sim$ 60). This behaviour suggests that the mass dependence of the fitting parameters is itself governed by a higher-order dependence on $Z$.

To further investigate this trend, the extracted intercepts and slopes are examined as functions of the proton number $Z$.

\begin{figure}[htbp]
\centering

\includegraphics[width=0.85\textwidth]{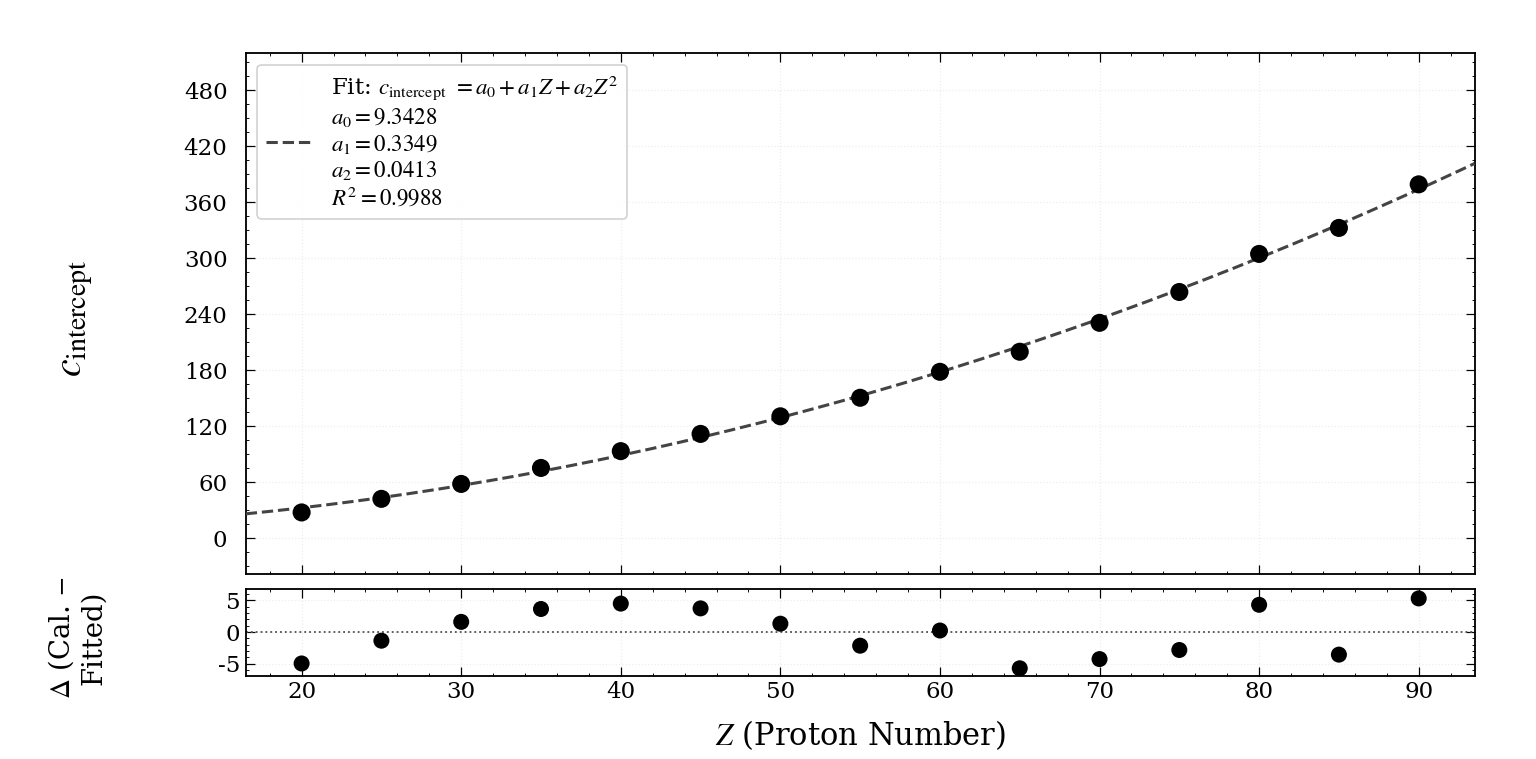}

\vspace{0.1cm}

\includegraphics[width=0.85\textwidth]{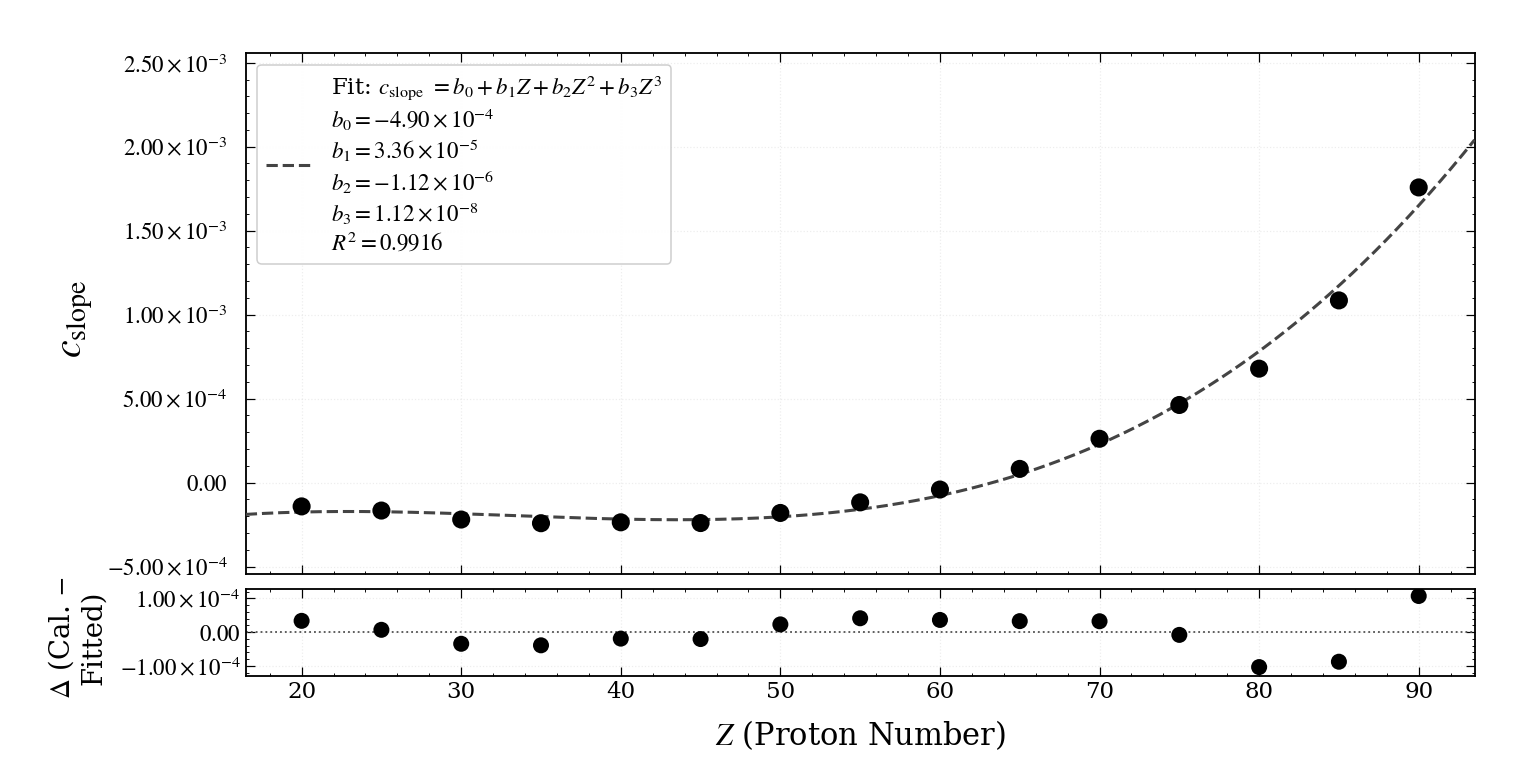}

\caption{Dependence of the fitting coefficients associated with parameter $c$ on proton number $Z$. The upper panel shows the variation of the intercept parameter $c_{\mathrm{intercept}}$, while the lower panel displays the corresponding slope parameter $c_{\mathrm{slope}}$. The solid curves represent the SR guided polynomial fits. The lower part of each of the plots indicates the difference between fitted and actual values.}
\label{fig:c_intercept_slope}
\end{figure}

\begin{figure}[htbp]
\centering

\includegraphics[width=1.0\textwidth]{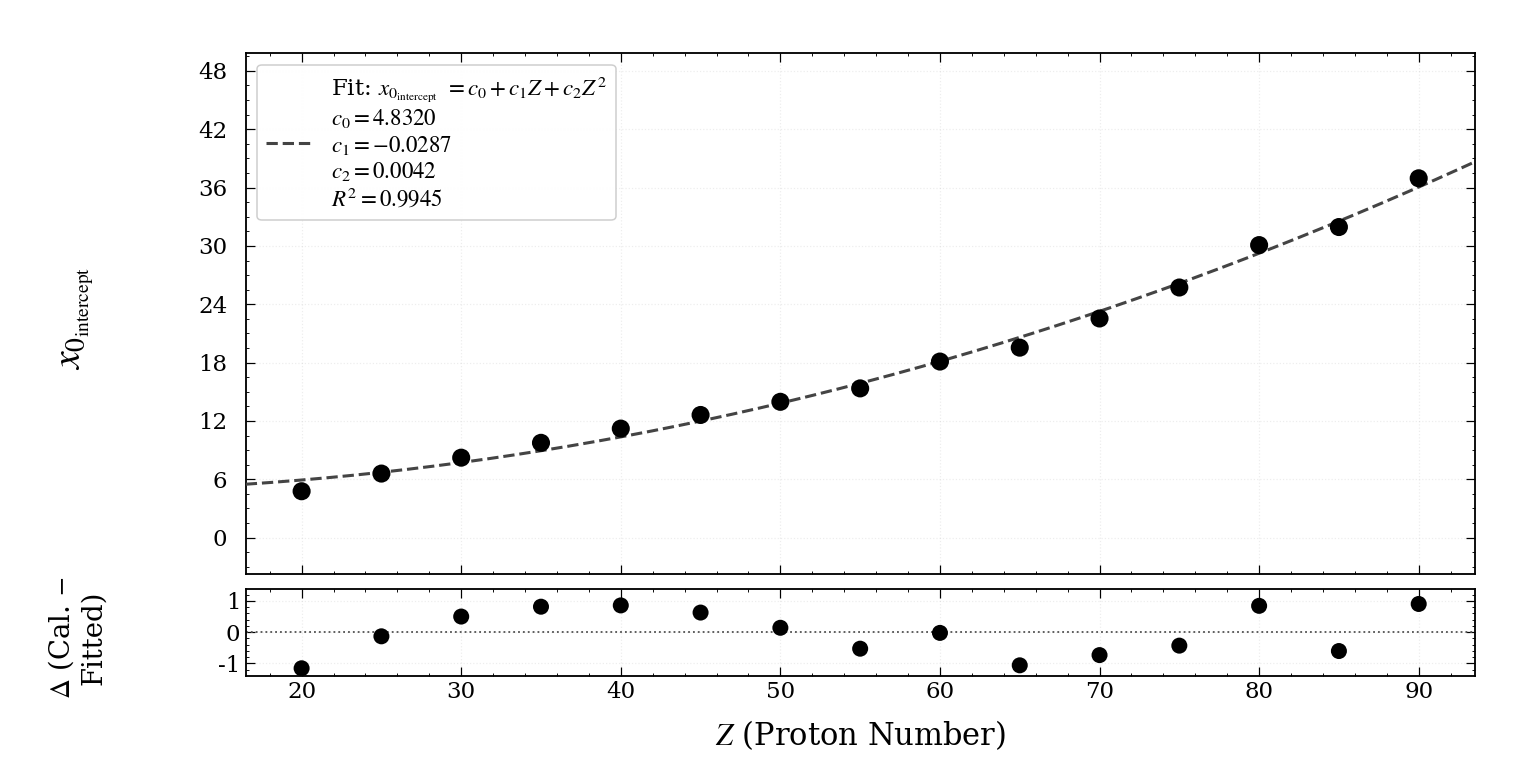}
\vspace{0.1cm}
\includegraphics[width=1.0\textwidth]{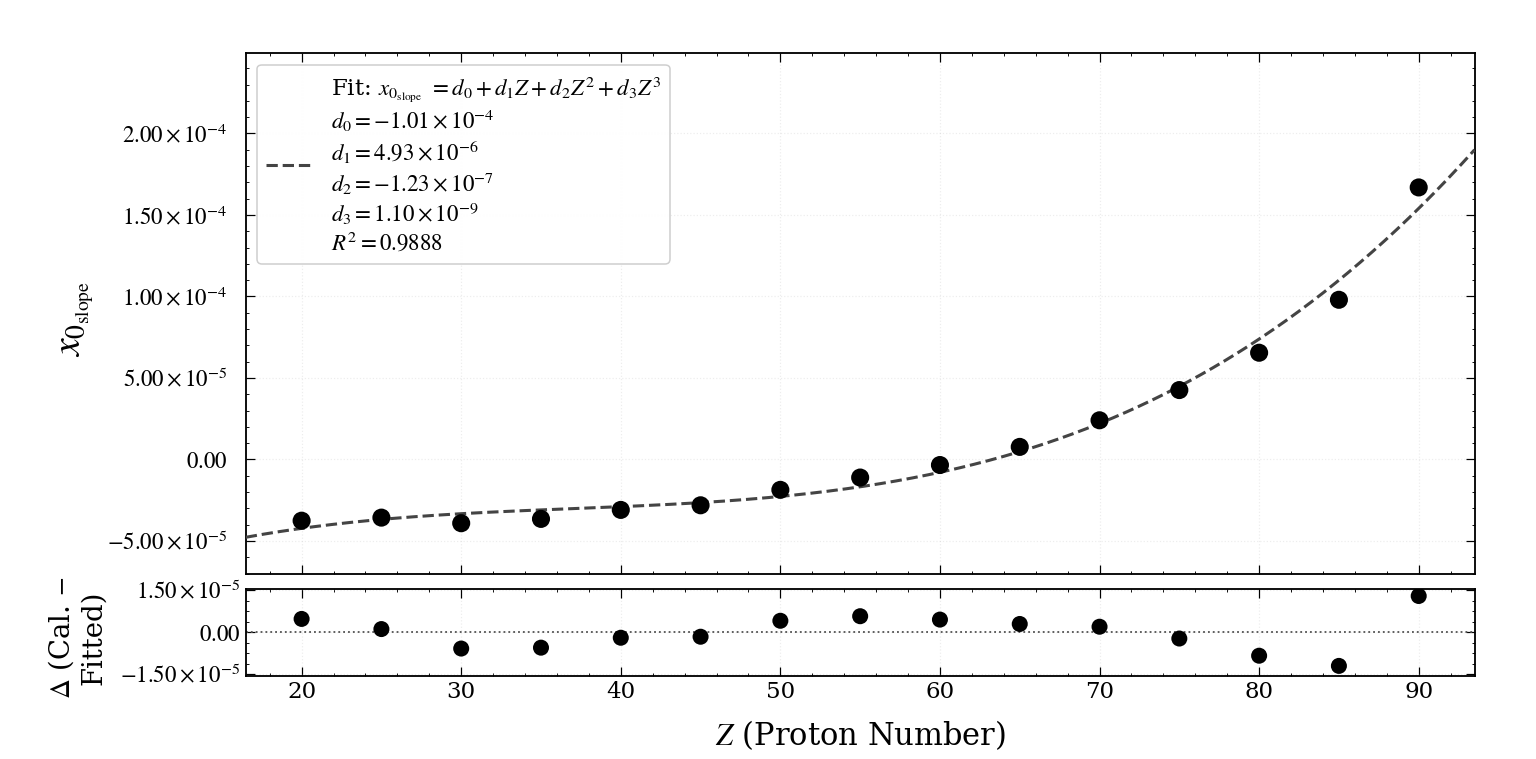}

\caption{Dependence of the fitting coefficients associated with parameter $x_0$ on proton number $Z$. The upper panel shows the variation of the intercept parameter $x_{0,\mathrm{intercept}}$, while the lower panel shows the slope parameter $x_{0,\mathrm{slope}}$. The curves correspond to the polynomial parametrisations obtained from SR. The lower part of each of the plots indicates the difference between fitted and actual values.}
\label{fig:x0_intercept_slope}
\end{figure}

The resulting trends are displayed in Figs.~\ref{fig:c_intercept_slope} and \ref{fig:x0_intercept_slope}. The machine learning guided SR analysis reveals that the intercept parameters, $c_{\mathrm{intercept}}$ and $x_{0,\mathrm{intercept}}$, are well described by quadratic functions of $Z$. In contrast, the slope parameters, $c_{\mathrm{slope}}$ and $x_{0,\mathrm{slope}}$, require cubic functions of $Z$ to accurately reproduce their behaviour. The different orders of functional dependence reflect the increasing complexity of the underlying nuclear structure effects as the proton number increases. Such effects may arise from the interplay of Coulomb interactions, shell evolution, and changes in the single particle structure across the nuclear chart.

\subsubsection{\textbf{Global Parametrisation}}

By combining the mass number and proton number dependencies discussed above, a global parametrisation for the decay rate ratio can be obtained. Substituting the $Z$ dependent forms of the intercept and slope parameters into the linear relations for $c(A)$ and $x_0(A)$ yields the following unified expression

\begin{equation}
\frac{\lambda_{bare}}{\lambda_{neutral}} = \exp\left(
\frac{
\left[a_0 + a_1 Z + a_2 Z^2\right] + A \left[b_0 + b_1 Z + b_2 Z^2 + b_3 Z^3\right]
}{
x + \left[c_0 + c_1 Z + c_2 Z^2\right] + A \left[d_0 + d_1 Z + d_2 Z^2 + d_3 Z^3\right]
}
\right),
\label{finaleqn}
\end{equation}

where $x \equiv Q_n$ denotes the neutral atom $\beta^-$ decay $Q$ value, and the coefficients ${a_i, b_i, c_i, d_i}$ are obtained from the machine learning guided regression analysis. Eq. ~\ref{finaleqn} thus provides a compact global representation of the decay rate ratio $\lambda_{bare}/\lambda_{neutral}$ in terms of the fundamental nuclear quantities $Z$, $A$, and $Q_n$. The resulting parametrisation enables rapid estimation of bare atom decay rates across a broad range of nuclei without the need for detailed microscopic calculations.

\begin{table}[htbp]
\centering
\caption{Optimized fitting parameters.}
\label{tab:fit_params}
\begin{tabular}{cc|cc}
\hline
\hline
Parameter & Value & Parameter & Value \\
\hline
$a_2$ & $4.126785\times10^{-2}$ & $c_2$ & $4.173646\times10^{-3}$ \\
$a_1$ & $3.348574\times10^{-1}$ & $c_1$ & $-2.871454\times10^{-2}$ \\
$a_0$ & $9.342763 \times10^{0}$              & $c_0$ & $4.831980 \times10^{0}$ \\
\hline
$b_3$ & $1.121273\times10^{-8}$ & $d_3$ & $1.102233\times10^{-9}$ \\
$b_2$ & $-1.118366\times10^{-6}$ & $d_2$ & $-1.225678\times10^{-7}$ \\
$b_1$ & $3.360850\times10^{-5}$ & $d_1$ & $4.933784\times10^{-6}$ \\
$b_0$ & $-4.901272\times10^{-4}$ & $d_0$ & $-1.007589\times10^{-4}$ \\
\hline
\end{tabular}
\end{table}


\paragraph{Validation of the Parametric Equation}

\begin{figure}[htbp]
\centering
\includegraphics[width=0.85\textwidth]{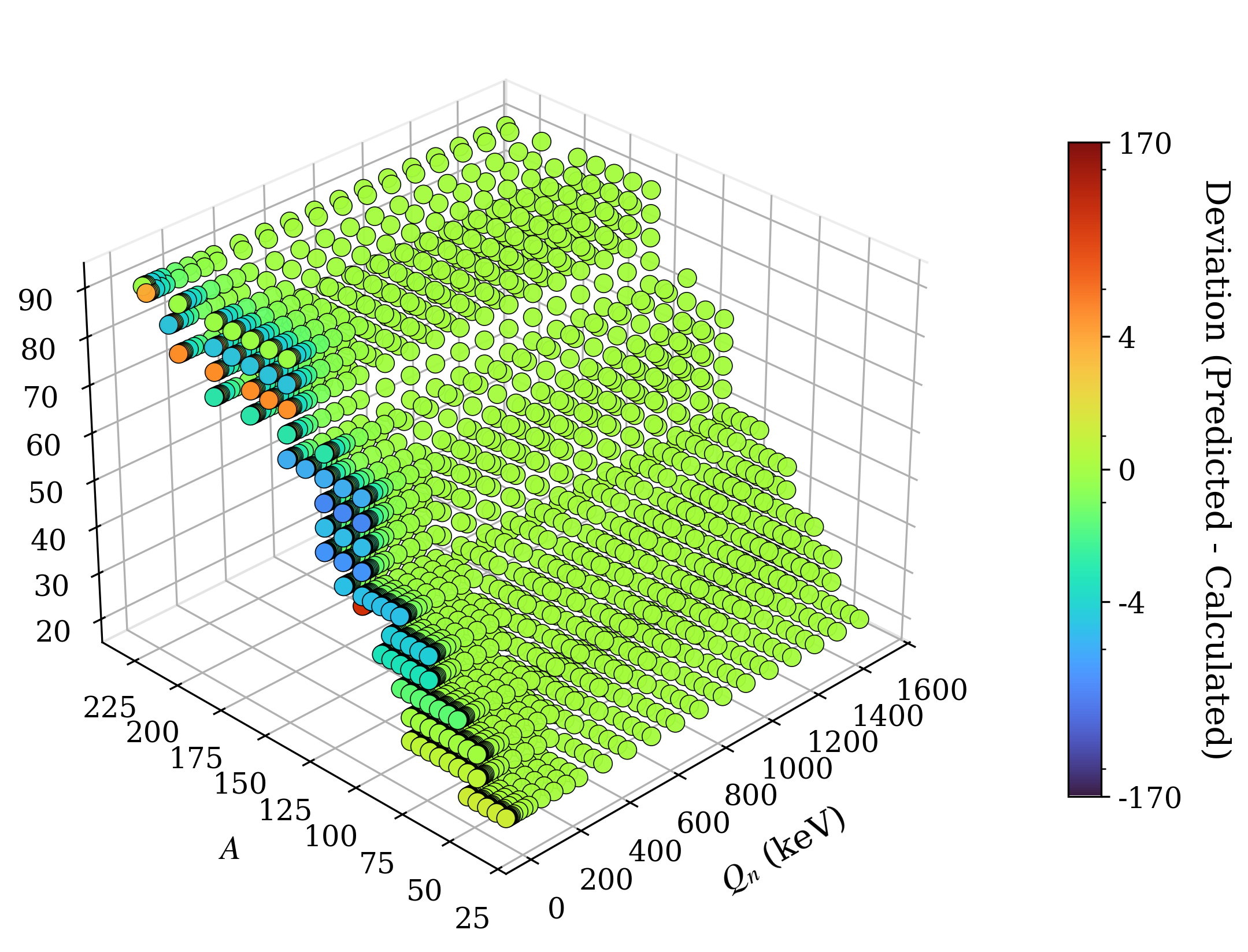}
\vspace{0.1cm}
\includegraphics[width=0.85\textwidth]{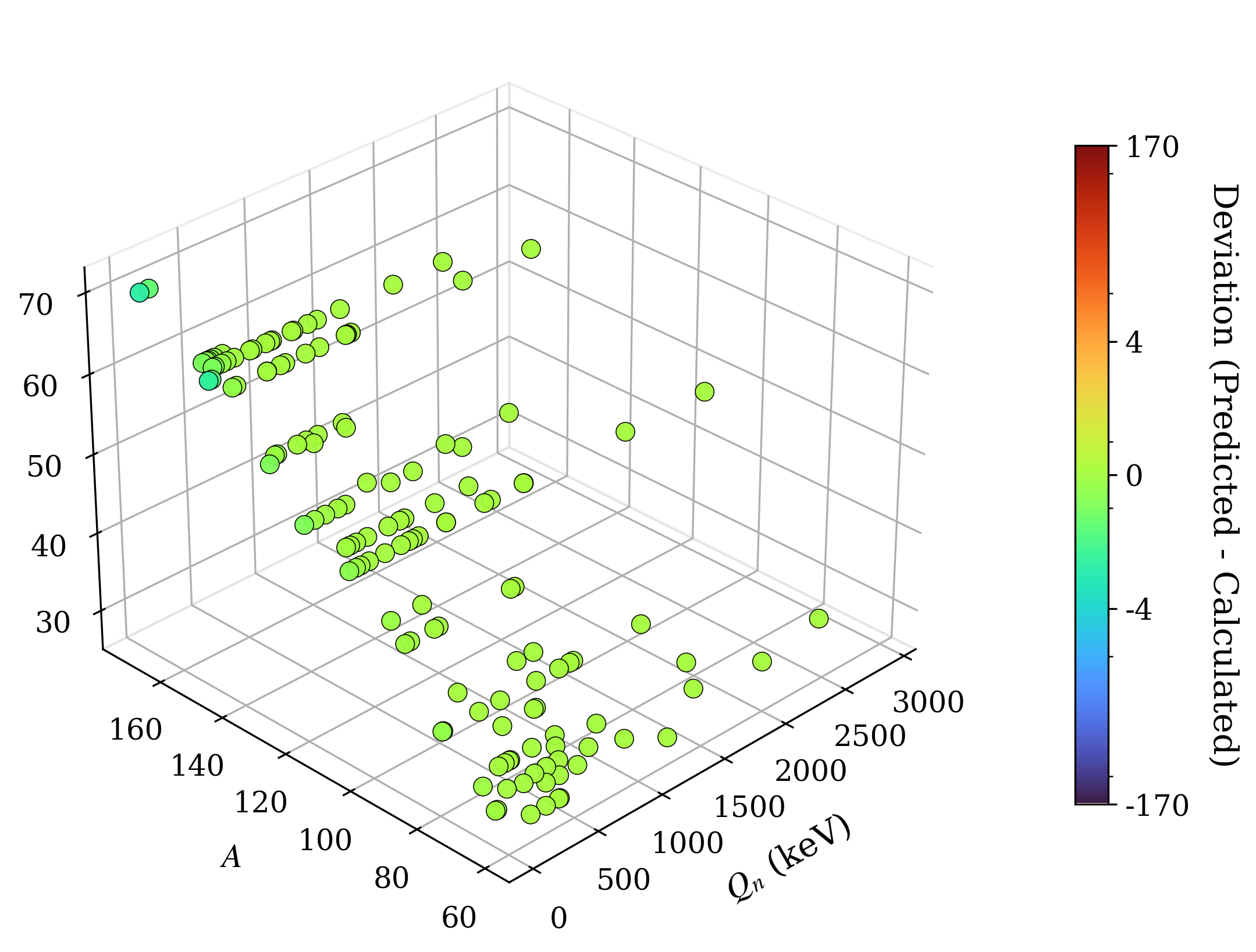}
\caption{Absolute deviation between the decay rate ratios predicted by Eq.~\ref{finaleqn} and the corresponding theoretical calculations as a function of the neutral atom decay energy $Q_n$, atomic number Z and mass number A. The upper panel shows the results for the nuclei of Set-1 used in constructing the parametrisation, while the lower panel presents the validation results for the independent Set-2 nuclei.}
\label{Formula}
\end{figure}

Finally, Eq.~\ref{finaleqn} was employed to predict the decay rate ratio of bare to neutral atoms for the nuclei belonging to Set-2. The predictive capability of the parametrisation was assessed by examining the absolute deviation between the calculated and predicted values as a function of the neutral atom decay energy, $Q_n$, as shown in Fig.~\ref{Formula}. The upper panel of Fig.~\ref{Formula} presents the deviations for the nuclei used in constructing the parametrisation (Set-1), while the lower panel shows the corresponding results for the independent validation dataset (Set-2).

For the Set-1 nuclei, the deviations remain small for the majority of transitions with relatively large decay energies ($Q_n > 150$ keV), irrespective of the atomic number $Z$ or mass number $A$. However, the magnitude and sign of the deviation exhibit a systematic dependence on $Z$, $A$, and $Q_n$. For low mass nuclei with small atomic numbers, the parametrisation tends to overestimate the $\lambda_{\mathrm{bare}}/\lambda_{\mathrm{neutral}}$ ratio when $Q_n < 100$ keV. In contrast, for medium  and high $Z$ nuclei, the parametrisation increasingly underestimates the decay rate ratio in the very low $Q_n$ region.  A similar behavior is observed for the independent Set-2 nuclei, indicating that the observed trends are systematic rather than dataset specific.

Overall, the close agreement between the predicted and calculated decay rate ratios demonstrates the robustness of the proposed parametrisation. The successful reproduction of the trends observed in both Set-1 and Set-2 provides confidence in its applicability to nuclei beyond the present dataset.

\subsection{\textbf{Machine Learning Prediction}}

\begin{figure}[htbp]
\centering
\includegraphics[width=0.85\textwidth]{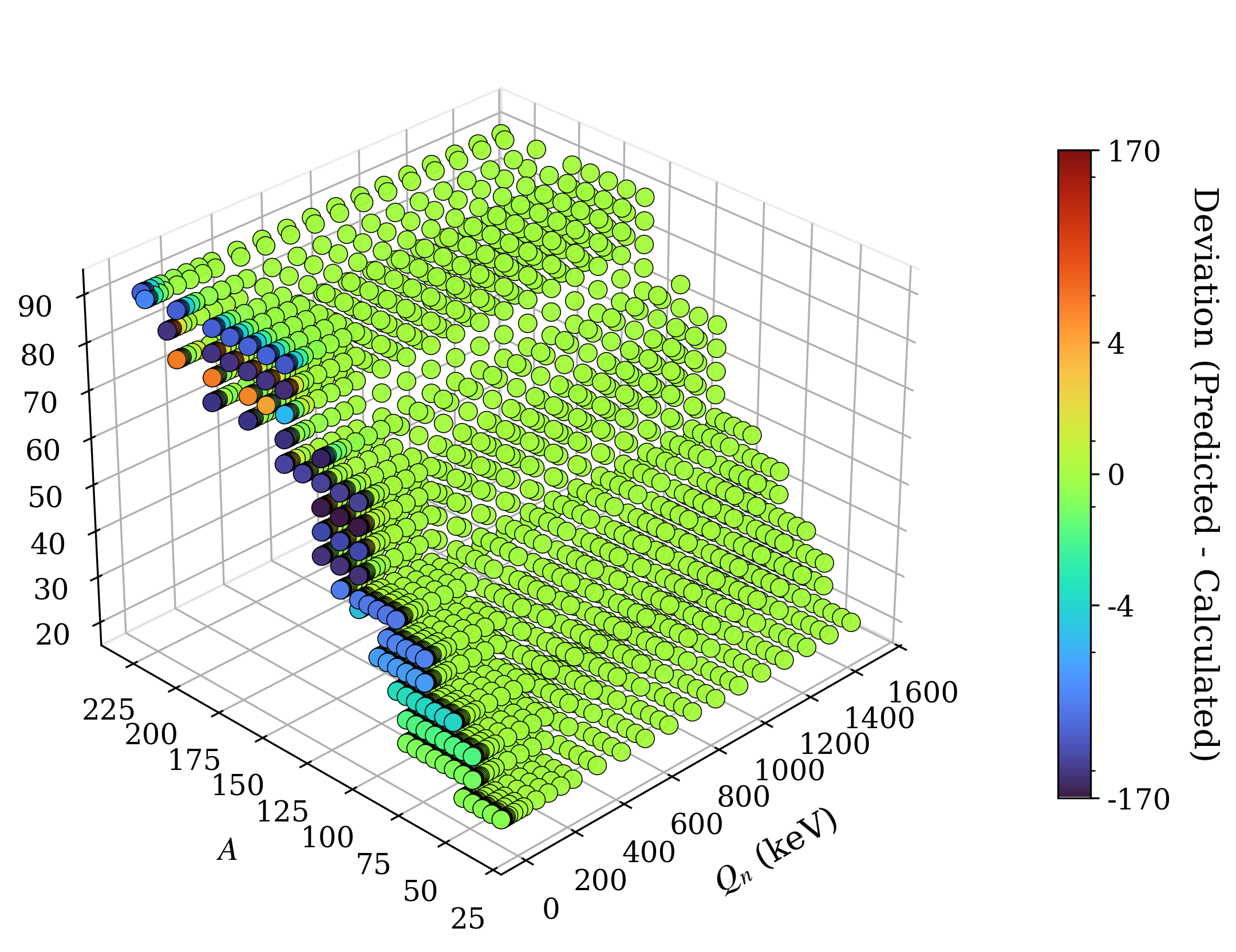}
\vspace{0.1cm}
\includegraphics[width=0.85\textwidth]{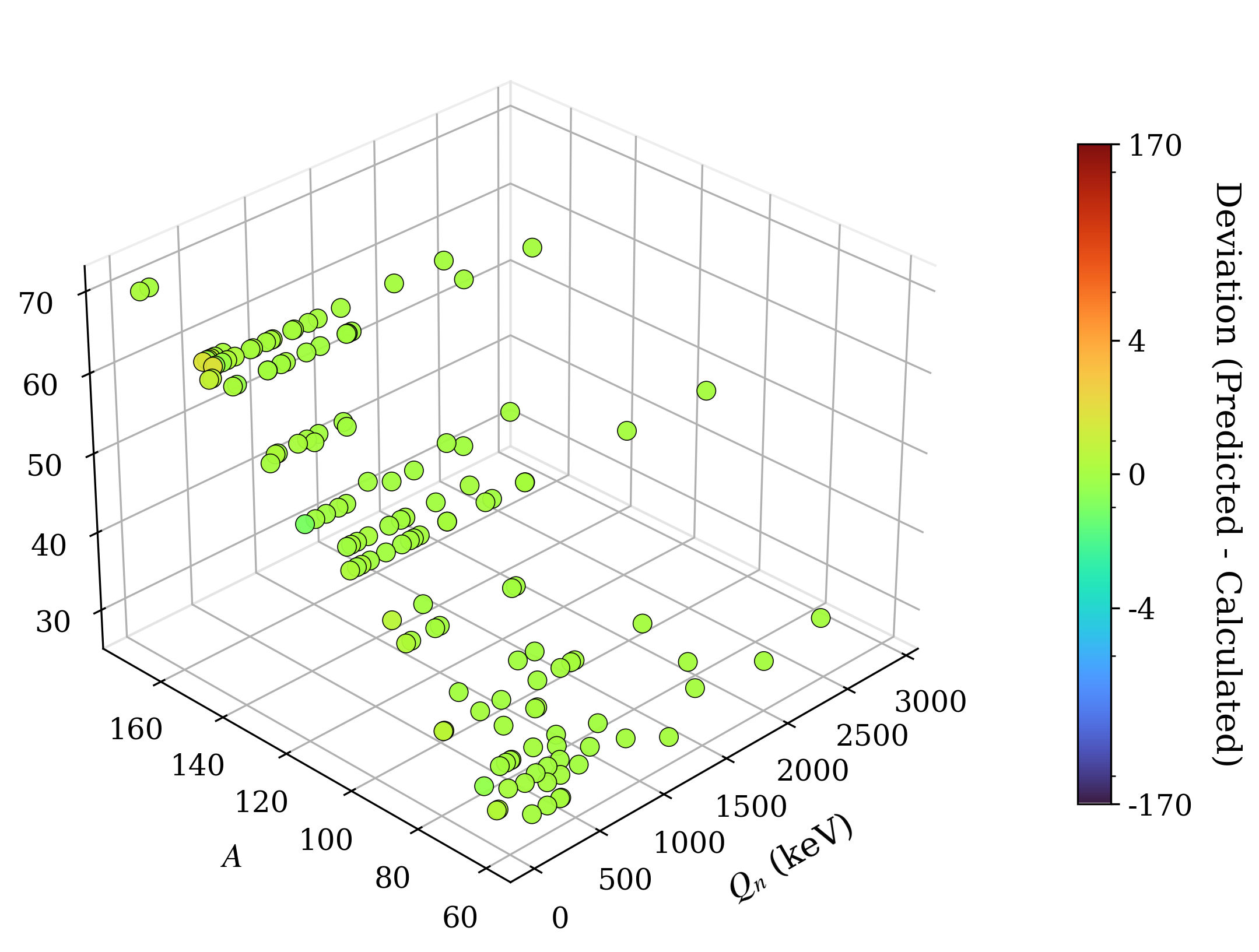}
\caption{Absolute deviation between the decay rate ratios predicted by RF model and the corresponding theoretical calculations as a function of the neutral atom decay energy $Q_n$, atomic number Z and mass number A. The upper panel shows the results for the nuclei of Set-1 used in training the model, while the lower panel presents the testing results for the independent Set-2 nuclei.}
\label{RF}
\end{figure}

\begin{figure}[htbp]
\centering
\includegraphics[width=0.85\textwidth]{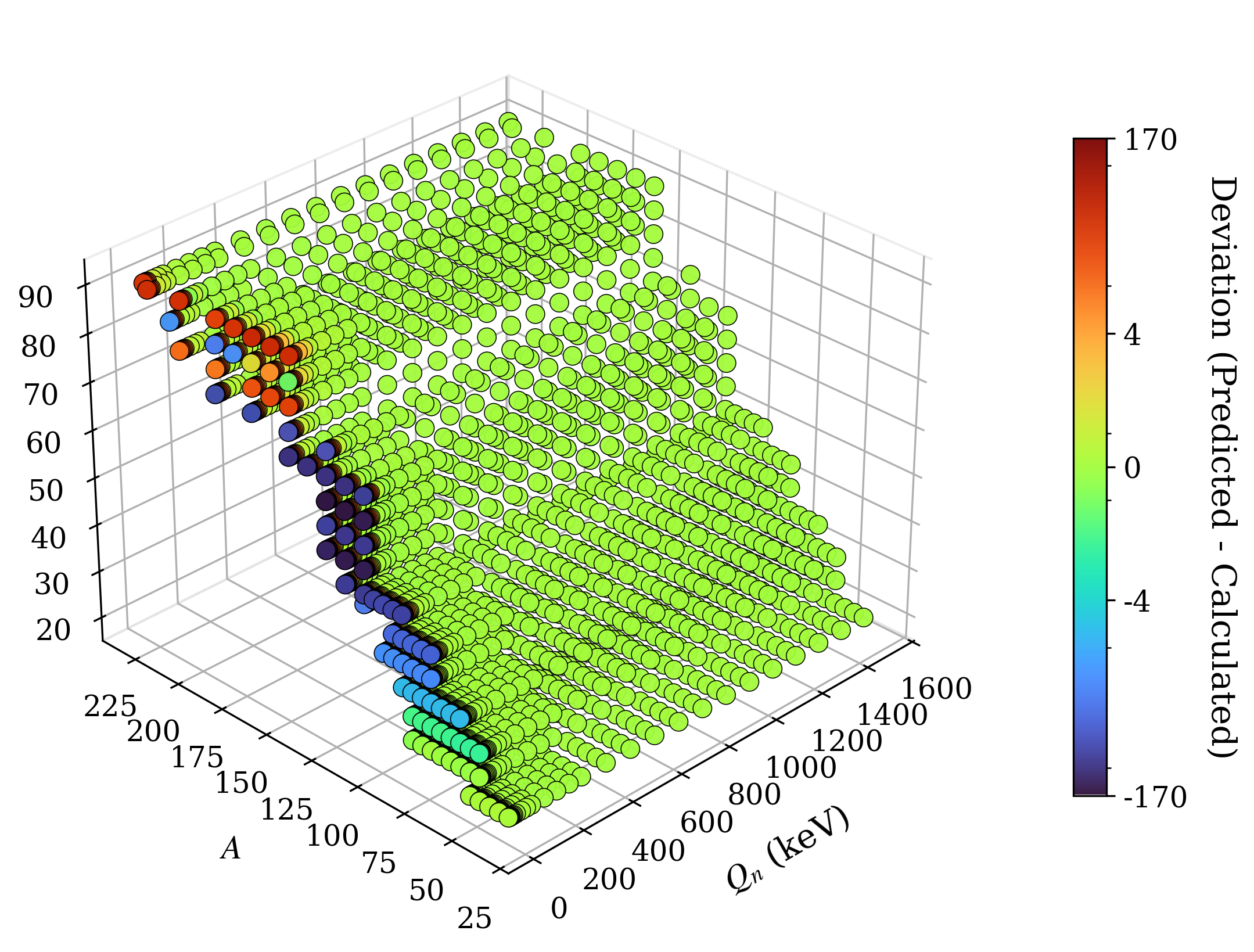}
\vspace{0.1cm}
\includegraphics[width=0.85\textwidth]{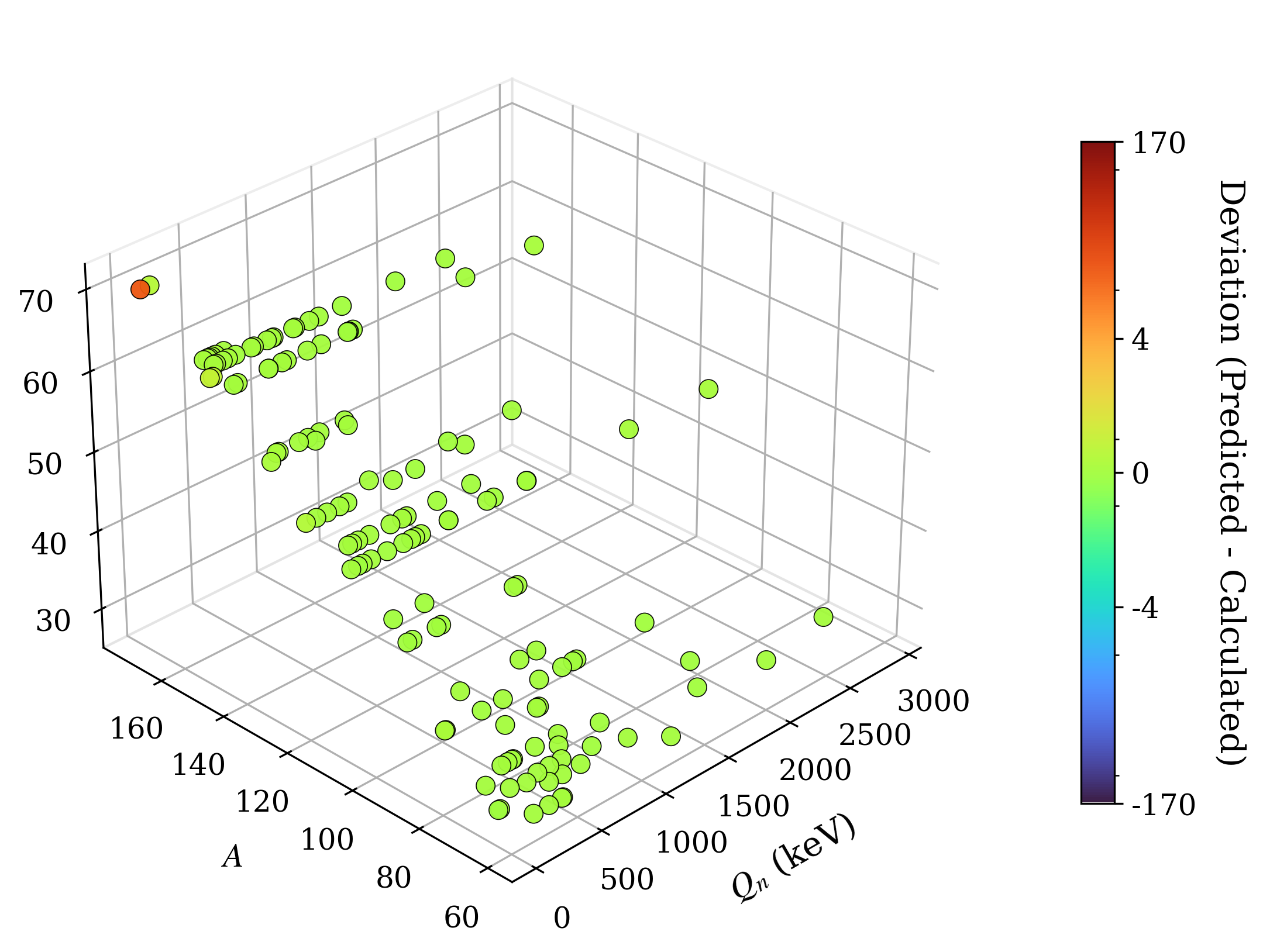}
\caption{Absolute deviation between the decay rate ratios predicted by ANN model and the corresponding theoretical calculations as a function of the neutral atom decay energy $Q_n$, atomic number Z and mass number A. The upper panel shows the results for the nuclei of Set-1 used in training the model, while the lower panel presents the testing results for the independent Set-2 nuclei.}
\label{NN}
\end{figure}

The predictive performance of the Artificial Neural Network (ANN) and Random Forest (RF) models was evaluated using the train data set (Set-1) and independent test dataset (Set-2). Figs. ~\ref{RF} and \ref{NN} compare the predicted values of $\lambda_{\mathrm{bare}}/\lambda_{\mathrm{neutral}}$ with the corresponding calculated values for both the training (upper panel) and test (lower panel) datasets. 

The Random Forest model accurately reproduces the overall trend of the data across both high and low $Q_{n}$ transitions. By 
utilizing an unrestricted feature space, the RF model effectively captures the nonlinear relationships for low $Q_{n}$ transitions across all intermediate and high Z elements and their isotopes. This robust predictive capability is consistently observed in both the training data and the independent test data set (Set-2).

Similarly, the ANN model exhibits significantly improved predictive performance. For the test dataset, the model reproduces the decay rate ratios with high accuracy, including the low $Q_n$ transitions of low $Z$ nuclei. As the atomic number increases to the intermediate $Z$ region, the predictive accuracy reduces slightly, and the model tends to underestimate the decay rate ratios. However, for high $Z$ nuclei, the trend reverses, with the model showing a tendency to overestimate the decay rate ratios for low $Q_n$ transitions. A similar systematic behavior is observed for the independent validation dataset (Set-2), indicating that these trends are intrinsic to the model rather than specific to the training data.


\subsection{\textbf{Performance Comparison of the Predictive Models}}

\begin{figure}[hbtp]
\centering
\includegraphics[width=1.0\textwidth]{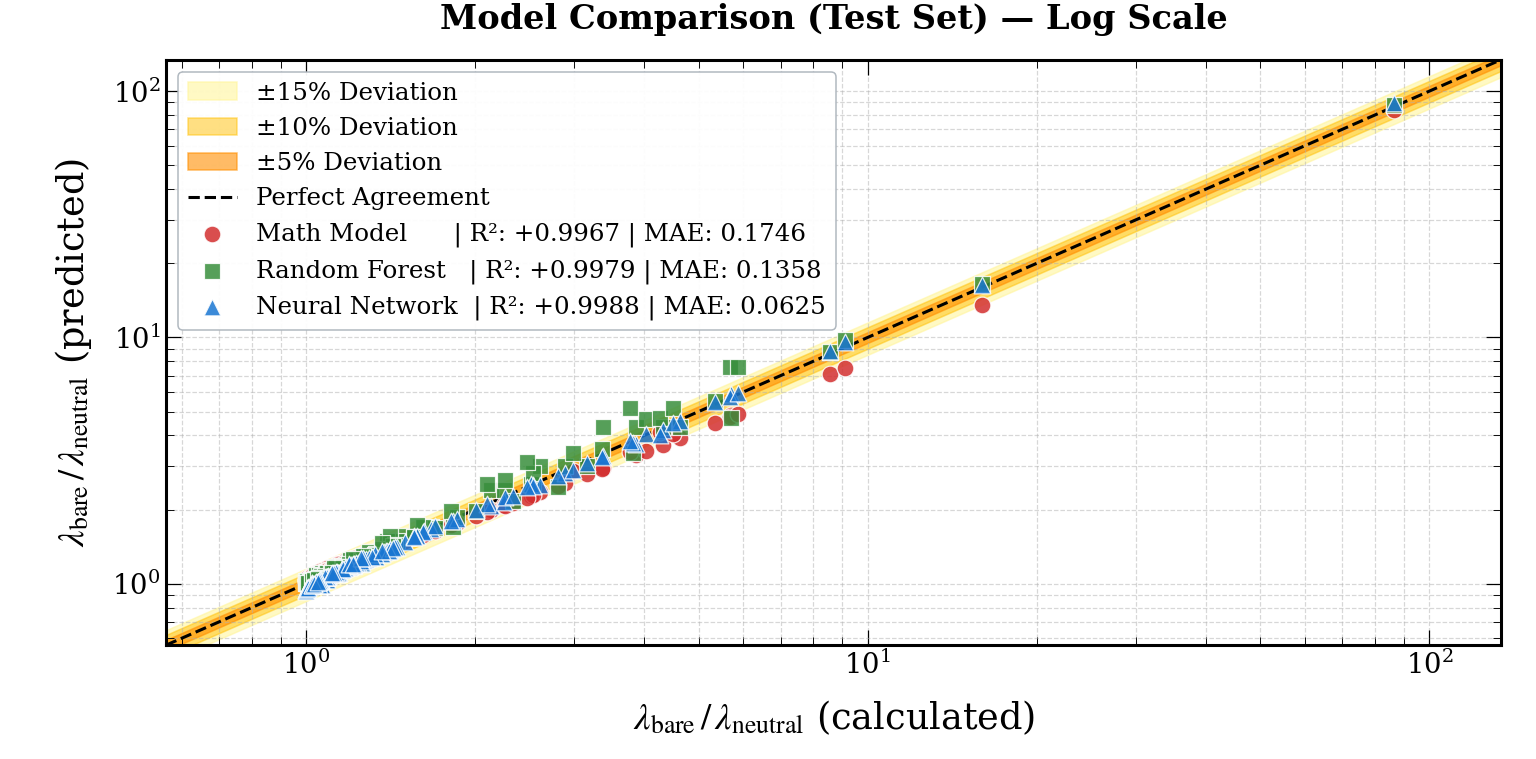}
\caption{Comparison of the predictive performance of the parametric, RF, and ANN models. The dashed diagonal line represents perfect agreement between the predicted and calculated values, while the shaded regions correspond to deviations of $\pm 5\%$, $\pm 10\%$, and $\pm 15\%$.}
\label{comparison}
\end{figure}

Fig. \ref{comparison} compares the predictive performance of the proposed parametric equation with those of the RF and ANN models for data Set-2. The dashed diagonal line represents perfect agreement between the predicted and calculated values, while the shaded bands indicate deviations of $\pm 5\%$, $\pm 10\%$, and $\pm 15\%$.

For the parametric formula, Fig.~\ref{comparison} shows that nearly $65\%$ of the data points lie within a deviation of $\pm 5\%$ from the calculated values, while approximately $25\%$ of the transitions are reproduced within $\pm 10\%$. In most cases, the deviation remains within $10\%$, although it reaches up to $15\%$ for a few transitions. The largest deviations are observed around $Q_n \approx 100$ keV. In contrast, for transitions with higher $Q_n$ values, the parametrisation tends to slightly overpredict the decay rate ratio, with the deviation typically remaining below $5\%$. When analyzed as a function of the mass number $A$, the sign of the deviation appears largely random. For low- and medium-mass nuclei ($A < 100$), the deviations generally remain within the range of $+5\%$ to $-10\%$. For heavier nuclei, however, deviations as large as $-15\%$ are observed for a few transitions. A similar behavior is found with respect to the atomic number $Z$. Low- and medium-$Z$ nuclei ($Z < 50$) exhibit comparatively smaller deviations, typically within $+5\%$ to $-10\%$. Quantitatively, the coefficient of determination is $R^2 = 0.9967$, indicating that the proposed parametrisation reproduces approximately $99.67\%$ of the variance present in the calculated $\lambda_{\mathrm{bare}}/\lambda_{\mathrm{neutral}}$ values. The remaining unexplained variance is less than $0.33\%$, demonstrating that the parametrized expression accurately captures the underlying dependence of the decay rate ratio over the entire dataset. Furthermore, the mean absolute error (MAE) is only $0.17$, implying that the predicted values deviate from the calculated values by less than $0.2$ units on average. Considering the broad range of $\lambda_{\mathrm{bare}}/\lambda_{\mathrm{neutral}}$ values covered by the data Set-2, this small error confirms the predictive accuracy and robustness of the proposed parametrisation.

To assess whether machine learning techniques can further improve the predictive accuracy, the same dataset was analyzed using RF and ANN models. The RF model reproduces the overall trend of the data with exceptional accuracy, achieving $R^{2}=0.9979$ and a mean absolute error $MAE=0.14$. Most data points remain tightly clustered around the perfect agreement line across the entire range of {}$ \lambda_{bare}/\lambda_{neutral}$ values, including {}low-$Q_{n}$ transitions. This performance indicates that 
the unconstrained decision tree structure of the RF model is highly effective in representing the smooth and strongly nonlinear dependence of the decay rate ratio on the input variables.

A substantial improvement is obtained with the neural network model, which demonstrates excellent predictive capability. For the test dataset, almost all points lie within the $\pm5\%$ deviation band and closely follow the perfect agreement line over the entire range of decay rate ratios. Quantitatively, the model achieves a coefficient of determination of $R^2 = 0.9988$ with only $\mathrm{MAE}=0.06$, confirming that the model successfully captures the underlying nonlinear relationship between the nuclear parameters $(Q_n, A, Z)$ and the decay rate ratio.

A closer examination of the deviations in the low $Q_n$ region reveals that the largest errors for all three models are concentrated in the region of heavier nuclei and extremely small $\beta^-$ decay energy releases (Figs. \ref{Formula}, \ref{RF} and \ref{NN}). For the parametric formula, for nuclei with $Z\sim48$, the percentage error reaches approximately $24\%$ at $Q_n\sim10$--$12$ keV, corresponding to an absolute error of about $35$. However, the error decreases sharply with increasing $Q_n$, with the absolute error falling below $1$ for $Q_n\sim25$ keV. A similar behaviour is observed for the RF model in approximately the same region of nuclei, where the prediction deviates from the calculated values by up to about $25\%$, with an absolute error of approximately $145$ at the smallest $Q_n$ values. The absolute error decreases to below $4$ for $Q_n\sim30$ keV, falls below $0.9$ around $Q_n\sim60$ keV, and becomes of the order of $0.03$ at $Q_n\sim200$ keV. The ANN model exhibits a comparable concentration of the maximum error in the same low $Q_n$ and intermediate mass region, with the largest deviation occurring around $A\sim120$, $Z\sim65$, where the absolute error is approximately $166$ and the percentage error reaches about $30\%$ at $Q_n\sim12$ keV. The absolute error decreases rapidly with increasing $Q_n$ and becomes approximately $2$ around $Q_n\sim30$ keV. These observations indicate that the absolute prediction error is strongly dependent on the location in the $(Q_n,A,Z)$ parameter space, with the most pronounced deviations occurring for extremely small $Q_n$ values in heavier nuclei, while the prediction accuracy improves rapidly as $Q_n$ increases.

A direct comparison of the three approaches highlights the relative strengths and limitations of each model. The ANN predictions remain tightly clustered around the perfect agreement line throughout the full parameter range. RF model performs on par with the ANN and exceeds the accuracy of the parametric equation.

Although the ANN and RF provides better predictive accuracy, the symbolic regression parametrisation retains important practical advantages. The SR approach provides an explicit analytical expression that is physically interpretable and computationally convenient, whereas the ANN and RF acts as a black box predictor. Consequently, the two approaches should be regarded as complementary rather than competing methodologies. The symbolic regression parametrisation offers physical insight and rapid analytical estimates, while the ANN and RF provides the highest predictive accuracy for practical calculations. The excellent agreement obtained for the independent test dataset demonstrates that machine learning methods can reliably predict the enhancement of $\beta^{-}$ decay rates in fully ionised atoms using only the nuclear quantities $Q_n$, $A$, and $Z$.

\subsection{\textbf{Prediction of Bare-to-Neutral Decay Rate Ratios}}

The analytical parametrisation, RF model, and ANN model developed in this work were employed to predict the $\lambda_{\mathrm{bare}}/\lambda_{\mathrm{neutral}}$ ratios for the nuclei included in data Set-3 and presented in table \ref{table 2}. Here, the spin-parity of the parent and daughter states are mentioned only for the allowed transitions of these nuclei. Forbidden transitions are beyond the scope of the present work. Furthermore, transitions involving uncertain spin-parity assignments, for which it is not possible to unambiguously determine whether the decay is allowed or forbidden, have been excluded from the dataset.

The three approaches exhibit a consistent overall trend, with the predicted enhancement depending primarily on the decay $Q_n$ value. For transitions with relatively large $Q_n$ values, the predicted ratio $\lambda_{\mathrm{bare}}/\lambda_{\mathrm{neutral}}$ remains close to unity, indicating that the contribution of bound state $\beta^-$ decay is small and the decay is dominated by the continuum channel. In contrast, the enhancement increases progressively with decreasing $Q_n$, reflecting the increasing contribution of the bound state decay channel. Consequently, nuclei with very low $Q_n$ values are predicted to exhibit the largest enhancement factors and therefore constitute the most promising candidates for future experimental investigations.

Overall, the predictions obtained from the analytical parametrisation, RF model, and ANN model show good agreement across most of the dataset. Minor differences are observed mainly for transitions with predicted ratios close to unity. In particular, the ANN model occasionally predicts values slightly below unity (typically in the range of 0.93–0.99). Since the ANN was trained as an unconstrained regression model without enforcing the physical condition $\lambda_{\mathrm{bare}}/\lambda_{\mathrm{neutral}} \geq 1$ \cite{Gupta2019}, such small deviations are attributed to regression uncertainty rather than a physically meaningful reduction in the decay rate. These deviations are confined to high $Q_n$ transitions, where the expected enhancement is intrinsically small, and therefore do not affect the identification of nuclei expected to exhibit significant bound-state enhancement.

The predicted $\lambda_{\mathrm{bare}}/\lambda_{\mathrm{neutral}}$ ratios presented for Set-3 provide estimates of the maximum possible enhancement of the $\beta^-$ decay rate in fully ionised atoms. These predictions may serve as valuable inputs for astrophysical nucleosynthesis calculations and as a guide for selecting promising candidates for future storage ring experiments aimed to investigate bound-state $\beta^-$ decay.

\begin{table}[htbp]
\centering
\caption{Comparison of the predicted $\lambda_{bare}/\lambda_{neutral}$ ratio obtained from the parametric equation, Random Forest (RF), and Artificial Neural Network (ANN) models for a set of nuclei having astrophysical importance. Here, [$E_i, J^{\pi}_i$] and [$E_f, J^{\pi}_f$] are the energy and spin parity of the parent and daughter states, respectively.} 
\label{table 2}
\resizebox{\textwidth}{!}{%
\begin{tabular}{cccccccc}
\hline
&&&&&&&\\
Parent & Daughter & [$E_i, J^{\pi}_i$] & [$E_f, J^{\pi}_f$] & $Q_n$ (keV) & $\frac{\lambda_{bare}}{\lambda_{neutral}}|_{Par. Eqn.}$ & $\frac{\lambda_{bare}}{\lambda_{neutral}}|_{RF}$ & $\frac{\lambda_{bare}}{\lambda_{neutral}}|_{ANN}$\\
&&&&&&&\\
\hline
\hline
$^{59}$Fe & $^{59}$Co & $[0,\;3/2^-]$ & $[1481.6,\;5/2^-]$ & 83.4 & 1.71 & 1.59 & 1.68 \\
$^{59}$Fe & $^{59}$Co & $[0,\;3/2^-]$ & $[1434.3,\;1/2^-]$ & 130.7 & 1.42 & 1.27 & 1.37 \\
$^{59}$Fe & $^{59}$Co & $[0,\;3/2^-]$ & $[1291.6,\;3/2^-]$ & 273.4 & 1.19 & 1.13 & 1.13 \\
$^{59}$Fe & $^{59}$Co & $[0,\;3/2^-]$ & $[1099.3,\;3/2^-]$ & 465.7 & 1.11 & 1.04 & 1.05 \\

\hline
$^{60}$Co & $^{60}$Ni & $[0,\;5^{+}]$ & $[2505.7,\;4^{+}]$ & 317.1 & 1.17 & 1.17 & 1.11 \\
\hline
\hline
\end{tabular}%
}
\end{table}


\begin{table}[htbp]
\addtocounter{table}{-1}
\centering
\caption{(Continued.)}
\resizebox{\textwidth}{!}{%
\begin{tabular}{cccccccc}
\hline
&&&&&&&\\
Parent & Daughter & [$E_i, J^{\pi}_i$] & [$E_f, J^{\pi}_f$] & $Q_n$ (keV) & $\frac{\lambda_{bare}}{\lambda_{neutral}}|_{Par. Eqn.}$ & $\frac{\lambda_{bare}}{\lambda_{neutral}}|_{RF}$ & $\frac{\lambda_{bare}}{\lambda_{neutral}}|_{ANN}$\\
&&&&&&&\\
\hline
\hline
$^{72}$Zn & $^{72}$Ga & $[0,\;0^{+}]$ & $[207.9,\;1^{+}]$ & 234.9 & 1.28 & 1.21 & 1.28 \\
$^{72}$Zn & $^{72}$Ga & $[0,\;0^{+}]$ & $[161.1,\;1^{+}]$ & 281.7 & 1.23 & 1.17 & 1.20 \\
$^{72}$Zn & $^{72}$Ga & $[0,\;0^{+}]$ & $[128.5,\;1,\;2]$ & 314.3 & 1.20 & 1.17 & 1.15 \\
$^{72}$Zn & $^{72}$Ga & $[0,\;0^{+}]$ & $[119.2,\;(0^{+})]$ & 323.6 & 1.20 & 1.17 & 1.14 \\
\hline



$^{72}$Ga & $^{72}$Ge & $[0,\; 3^-]$ & $[3816,\;2^-,\;3^-]$ & 181.5 & 1.39 & 1.29 & 1.43 \\
\hline
\hline
\end{tabular}%
}
\end{table}


\begin{table}[htbp]
\addtocounter{table}{-1}
\centering
\caption{(Continued.)}
\resizebox{\textwidth}{!}{%
\begin{tabular}{cccccccc}
\hline
&&&&&&&\\
Parent & Daughter & [$E_i, J^{\pi}_i$] & [$E_f, J^{\pi}_f$] & $Q_n$ (keV) & $\frac{\lambda_{bare}}{\lambda_{neutral}}|_{Par. Eqn.}$ & $\frac{\lambda_{bare}}{\lambda_{neutral}}|_{RF}$ & $\frac{\lambda_{bare}}{\lambda_{neutral}}|_{ANN}$\\
&&&&&&&\\
\hline
\hline
$^{78}$Ge & $^{78}$As & $[0,\;0^{+}]$ & $[293.9,\;1^{+}]$ & 661.1 & 1.10 & 1.06 & 1.03 \\

$^{78}$Ge & $^{78}$As & $[0,\;0^{+}]$ & $[277.3,\;1^{+}]$ & 677.7 & 1.10 & 1.06 & 1.02 \\

\hline

$^{78}$As & $^{78}$Se & $[0,\; 2^-]$ & $[3411.5,\; 3^-]$ & 797.5 & 1.09 & 1.04 & 1.01 \\

$^{78}$As & $^{78}$Se & $[0,\; 2^-]$ & $[3372.7,\; 3^-]$ & 836.3 & 1.08 & 1.04 & 1.00 \\

$^{78}$As & $^{78}$Se & $[0,\; 2^-]$ & $[3144.5,\; 3^-]$ & 1064.5 & 1.07 & 1.02 & 0.98 \\
\hline
\hline
\end{tabular}%
}
\end{table}


\begin{table}[htbp]
\addtocounter{table}{-1}
\centering
\caption{(Continued.)}
\resizebox{\textwidth}{!}{%
\begin{tabular}{cccccccc}
\hline
&&&&&&&\\
Parent & Daughter & [$E_i, J^{\pi}_i$] & [$E_f, J^{\pi}_f$] & $Q_n$ (keV) & $\frac{\lambda_{bare}}{\lambda_{neutral}}|_{Par. Eqn.}$ & $\frac{\lambda_{bare}}{\lambda_{neutral}}|_{RF}$ & $\frac{\lambda_{bare}}{\lambda_{neutral}}|_{ANN}$\\
&&&&&&&\\
\hline
\hline
$^{96}$Nb & $^{96}$Mo & $[0,\;6^{+}]$ & $[2975.1,\;5^{+}]$ & 211.9 & 1.54 & 1.55 & 1.60 \\
$^{96}$Nb & $^{96}$Mo & $[0,\;6^{+}]$ & $[2875.5,\;7^{+}]$ & 311.5 & 1.35 & 1.34 & 1.32 \\
$^{96}$Nb & $^{96}$Mo & $[0,\;6^{+}]$ & $[2755.2,\;6^{+}]$ & 431.8 & 1.24 & 1.21 & 1.18 \\
$^{96}$Nb & $^{96}$Mo & $[0,\;6^{+}]$ & $[2438.5,\;5^{+}]$ & 748.5 & 1.14 & 1.08 & 1.06 \\
$^{96}$Nb & $^{96}$Mo & $[0,\;6^{+}]$ & $[2440.8,\;6^{+}]$ & 746.2 & 1.14 & 1.08 & 1.06 \\
\hline
\hline
\end{tabular}%
}
\end{table}


\begin{table}[htbp]
\addtocounter{table}{-1}
\centering
\caption{(Continued.)}
\resizebox{\textwidth}{!}{%
\begin{tabular}{cccccccc}
\hline
&&&&&&&\\
Parent & Daughter & [$E_i, J^{\pi}_i$] & [$E_f, J^{\pi}_f$] & $Q_n$ (keV) & $\frac{\lambda_{bare}}{\lambda_{neutral}}|_{Par. Eqn.}$ & $\frac{\lambda_{bare}}{\lambda_{neutral}}|_{RF}$ & $\frac{\lambda_{bare}}{\lambda_{neutral}}|_{ANN}$\\
&&&&&&&\\
\hline
\hline
$^{106}$Rh & $^{106}$Pd & $[0,\;1^{+}]$ & $[3320.5,\;0^{+}]$ & 220.5 & 1.62 & 1.80 & 1.68 \\

$^{106}$Rh & $^{106}$Pd & $[0,\;1^{+}]$ & $[3252,\;2^{+}]$ & 289.0 & 1.45 & 1.48 & 1.42 \\

$^{106}$Rh & $^{106}$Pd & $[0,\;1^{+}]$ & $[3221.4,\;0^{+}]$ & 319.6 & 1.40 & 1.48 & 1.36 \\

$^{106}$Rh & $^{106}$Pd & $[0,\;1^{+}]$ & $[3054.9,\;1^{+}]$ & 486.1 & 1.25 & 1.20 & 1.19 \\

$^{106}$Rh & $^{106}$Pd & $[0,\;1^{+}]$ & $[2968.7,\;2^{+}]$ & 572.3 & 1.21 & 1.15 & 1.14 \\
\hline
\hline
\end{tabular}%
}
\end{table}


\begin{table}[htbp]
\addtocounter{table}{-1}
\centering
\caption{(Continued.)}
\resizebox{\textwidth}{!}{%
\begin{tabular}{cccccccc}
\hline
&&&&&&&\\
Parent & Daughter & [$E_i, J^{\pi}_i$] & [$E_f, J^{\pi}_f$] & $Q_n$ (keV) & $\frac{\lambda_{bare}}{\lambda_{neutral}}|_{Par. Eqn.}$ & $\frac{\lambda_{bare}}{\lambda_{neutral}}|_{RF}$ & $\frac{\lambda_{bare}}{\lambda_{neutral}}|_{ANN}$\\
&&&&&&&\\
\hline
\hline
$^{111}$Pd & $^{111}$Ag & $[0,\;5/2^{+}]$ & $[1704.8,\;(5/2^{+} 7/2^{+})]$ & 512.2 & 1.25 & 1.21 & 1.18 \\

$^{111}$Pd & $^{111}$Ag & $[0,\;5/2^{+}]$ & $[1621.9,\;3/2^{+}]$ & 595.1 & 1.21 & 1.15 & 1.14 \\

$^{111}$Pd & $^{111}$Ag & $[0,\;5/2^{+}]$ & $[1602.5,\;5/2^{+}]$ & 614.5 & 1.20 & 1.15 & 1.13 \\

$^{111}$Pd & $^{111}$Ag & $[0,\;5/2^{+}]$ & $[1518.8,\;5/2^{+} 7/2^{+}]$ & 698.2 & 1.18 & 1.12 & 1.10 \\

$^{111}$Pd & $^{111}$Ag & $[0,\;5/2^{+}]$ & $[1210.4,\;3/2^{+}]$ & 1006.6 & 1.12 & 1.06 & 1.02 \\
\hline
\hline
\end{tabular}%
}
\end{table}


\begin{table}[htbp]
\addtocounter{table}{-1}
\centering
\caption{(Continued.)}
\resizebox{\textwidth}{!}{%
\begin{tabular}{cccccccc}
\hline
&&&&&&&\\
Parent & Daughter & [$E_i, J^{\pi}_i$] & [$E_f, J^{\pi}_f$] & $Q_n$ (keV) & $\frac{\lambda_{bare}}{\lambda_{neutral}}|_{Par. Eqn.}$ & $\frac{\lambda_{bare}}{\lambda_{neutral}}|_{RF}$ & $\frac{\lambda_{bare}}{\lambda_{neutral}}|_{ANN}$\\
&&&&&&&\\
\hline
\hline
\hline$^{117}$In & $^{117}$Sn & $[0,\;9/2^{+}]$ & $[711.5,\;7/2^{+}]$ & 743.5 & 1.19 & 1.14 & 1.11 \\
\hline
$^{124}$Sb & $^{124}$Te & $[10.8627,\;5^{+}]$ & $[2349.5,\;6^{+}]$ & 565.7 & 1.27 & 1.18 & 1.20 \\

$^{124}$Sb & $^{124}$Te & $[10.8627,\;5^{+}]$ & $[1746.9,\;6^{+}]$ & 1168.3 & 1.12 & 1.05 & 1.02 \\

$^{124}$Sb & $^{124}$Te & $[10.8627,\;5^{+}]$ & $[1248.5,\;4^{+}]$ & 1666.7 & 1.09 & 1.03 & 0.99 \\

\hdashline
$^{124}$Sb & $^{124}$Te & $[0,\; 3^-]$ & $[2886.4,\; 3^-]$ & 17.9 & 70.33 & 63.93 & 85.42 \\
\hline
\hline
\end{tabular}%
}
\end{table}


\begin{table}[htbp]
\addtocounter{table}{-1}
\centering
\caption{(Continued.)}
\resizebox{\textwidth}{!}{%
\begin{tabular}{cccccccc}
\hline
&&&&&&&\\
Parent & Daughter & [$E_i, J^{\pi}_i$] & [$E_f, J^{\pi}_f$] & $Q_n$ (keV) & $\frac{\lambda_{bare}}{\lambda_{neutral}}|_{Par. Eqn.}$ & $\frac{\lambda_{bare}}{\lambda_{neutral}}|_{RF}$ & $\frac{\lambda_{bare}}{\lambda_{neutral}}|_{ANN}$\\
&&&&&&&\\ 

\hline
\hline

$^{129}$Sb & $^{129}$Te & $[0,\;7/2^{+}]$ & $[2265.3,\;(5/2^{+},\; 7/2^{+})]$ & 110.7 & 3.02 & 3.89 & 3.54 \\

$^{129}$Sb & $^{129}$Te & $[0,\;7/2^{+}]$ & $[2114.6,\;5/2^{+}]$ & 261.4 & 1.65 & 1.69 & 1.69 \\

$^{129}$Sb & $^{129}$Te & $[0,\;7/2^{+}]$ & $[2086.1,\;(7/2^{+})]$ & 289.9 & 1.57 & 1.56 & 1.58 \\

$^{129}$Sb & $^{129}$Te & $[0,\;7/2^{+}]$ & $[2071.4,\;5/2^{+}]$ & 304.6 & 1.54 & 1.56 & 1.55 \\

$^{129}$Sb & $^{129}$Te & $[0,\;7/2^{+}]$ & $[1921.3,\;(5/2)^{+}]$ & 454.7 & 1.34 & 1.24 & 1.26 \\
\hline
\hline
\end{tabular}%
}
\end{table}


\begin{table}[htbp]
\addtocounter{table}{-1}
\centering
\caption{(Continued.)}
\resizebox{\textwidth}{!}{%
\begin{tabular}{cccccccc}
\hline
&&&&&&&\\
Parent & Daughter & [$E_i, J^{\pi}_i$] & [$E_f, J^{\pi}_f$] & $Q_n$ (keV) & $\frac{\lambda_{bare}}{\lambda_{neutral}}|_{Par. Eqn.}$ & $\frac{\lambda_{bare}}{\lambda_{neutral}}|_{RF}$ & $\frac{\lambda_{bare}}{\lambda_{neutral}}|_{ANN}$\\
&&&&&&&\\
\hline
\hline
$^{131}$Te & $^{131}$I & $[0,\;3/2^{+}]$ & $[2072.6,\;3/2^{+},\; 5/2^{+}]$ & 162.3 & 2.24 & 2.79 & 2.45 \\

$^{131}$Te & $^{131}$I & $[0,\;3/2^{+}]$ & $[1800.6,\;3/2^{+},\; 5/2^{+}]$ & 434.3 & 1.37 & 1.43 & 1.30 \\

$^{131}$Te & $^{131}$I & $[0,\;3/2^{+}]$ & $[1500.6,\;3/2^{+},\; 5/2^{+}]$ & 734.3 & 1.21 & 1.18 & 1.13 \\

$^{131}$Te & $^{131}$I & $[0,\;3/2^{+}]$ & $[1427.2,\;3/2^{+},\; 5/2^{+}]$ & 807.7 & 1.19 & 1.14 & 1.10 \\

$^{131}$Te & $^{131}$I & $[0,\;3/2^{+}]$ & $[1346.5,\;1/2^{+},\; 3/2^{+}]$ & 888.4 & 1.17 & 1.12 & 1.08 \\
\hline
\hline
\end{tabular}%
}
\end{table}


\begin{table}[htbp]
\addtocounter{table}{-1}
\centering
\caption{(Continued.)}
\resizebox{\textwidth}{!}{%
\begin{tabular}{cccccccc}
\hline
&&&&&&&\\
Parent & Daughter & [$E_i, J^{\pi}_i$] & [$E_f, J^{\pi}_f$] & $Q_n$ (keV) & $\frac{\lambda_{bare}}{\lambda_{neutral}}|_{Par. Eqn.}$ & $\frac{\lambda_{bare}}{\lambda_{neutral}}|_{RF}$ & $\frac{\lambda_{bare}}{\lambda_{neutral}}|_{ANN}$\\
&&&&&&&\\
\hline
\hline
\hline$^{134}$I & $^{134}$Xe & $[316.49,\;(8)^-]$ & $[1966.3,\;7^-]$ & 2525.2 & 1.06 & 1.05 & 0.98 \\

\hdashline
$^{134}$I & $^{134}$Xe & $[0,\;(4)^{+}]$ & $[2867.4,\;4^{+}]$ & 1307.6 & 1.12 & 1.06 & 1.01 \\

$^{134}$I & $^{134}$Xe & $[0,\;(4)^{+}]$ & $[2653.9,\;3^{+}]$ & 1521.1 & 1.10 & 1.05 & 0.99 \\

$^{134}$I & $^{134}$Xe & $[0,\;(4)^{+}]$ & $[2588.5,\;4^{+}]$ & 1586.5 & 1.10 & 1.05 & 0.99 \\

$^{134}$I & $^{134}$Xe & $[0,\;(4)^{+}]$ & $[2547.6,\;4^{+},\; 5^{+}]$ & 1627.4 & 1.05 & 1.12 & 0.99 \\
\hline
\hline
\end{tabular}%
}
\end{table}

\begin{table}[htbp]
\addtocounter{table}{-1}
\centering
\caption{(Continued.)}
\resizebox{\textwidth}{!}{%
\begin{tabular}{cccccccc}
\hline
&&&&&&&\\
Parent & Daughter & [$E_i, J^{\pi}_i$] & [$E_f, J^{\pi}_f$] & $Q_n$ (keV) & $\frac{\lambda_{bare}}{\lambda_{neutral}}|_{Par. Eqn.}$ & $\frac{\lambda_{bare}}{\lambda_{neutral}}|_{RF}$ & $\frac{\lambda_{bare}}{\lambda_{neutral}}|_{ANN}$\\
&&&&&&&\\
\hline
\hline
$^{151}$Pm & $^{151}$Sm & $[0,\;5/2^{+}]$ & $[964.2,\;5/2(^{+})]$ & 222.8 & 2.18 & 2.46 & 2.26 \\

$^{151}$Pm & $^{151}$Sm & $[0,\;5/2^{+}]$ & $[953.5,\;3/2(^{+})]$ & 233.5 & 2.11 & 2.08 & 2.18 \\

$^{151}$Pm & $^{151}$Sm & $[0,\;5/2^{+}]$ & $[877.6,\;5/2(^{+})]$ & 309.4 & 1.78 & 1.92 & 1.82 \\

$^{151}$Pm & $^{151}$Sm & $[0,\;5/2^{+}]$ & $[822.6,\;(3/2^{+},\;5/2^{+})]$ & 364.4 & 1.64 & 1.54 & 1.65 \\

$^{151}$Pm & $^{151}$Sm & $[0,\;5/2^{+}]$ & $[774,\;5/2(^{+})]$ & 413.0 & 1.55 & 1.54 & 1.56 \\
\hline
\hline
\end{tabular}%
}
\end{table}

\begin{table}[htbp]
\addtocounter{table}{-1}
\centering
\caption{(Continued.)}
\resizebox{\textwidth}{!}{%
\begin{tabular}{cccccccc}
\hline
&&&&&&&\\
Parent & Daughter & [$E_i, J^{\pi}_i$] & [$E_f, J^{\pi}_f$] & $Q_n$ (keV) & $\frac{\lambda_{bare}}{\lambda_{neutral}}|_{Par. Eqn.}$ & $\frac{\lambda_{bare}}{\lambda_{neutral}}|_{RF}$ & $\frac{\lambda_{bare}}{\lambda_{neutral}}|_{ANN}$\\
&&&&&&&\\
\hline
\hline

$^{165}$Dy & $^{165}$Ho & $[0,\;7/2^{+}]$ & $[1186.6,\;9/2^{+}]$ & 99.1 & 6.03 & 7.66 & 7.47 \\

$^{165}$Dy & $^{165}$Ho & $[0,\;7/2^{+}]$ & $[1079.6,\;7/2^{+}]$ & 206.1 & 2.59 & 3.03 & 2.81 \\

$^{165}$Dy & $^{165}$Ho & $[0,\;7/2^{+}]$ & $[1055.8,\;5/2^{+}]$ & 229.9 & 2.37 & 2.40 & 2.48 \\

$^{165}$Dy & $^{165}$Ho & $[0,\;7/2^{+}]$ & $[995.1,\;5/2^{+}]$ & 290.6 & 2.00 & 2.13 & 2.04 \\

$^{165}$Dy & $^{165}$Ho & $[0,\;7/2^{+}]$ & $[820.1,\;(9/2^{+})]$ & 465.6 & 1.56 & 1.49 & 1.57 \\
\hline
\hline
\end{tabular}%
}
\end{table}
\begin{table}[htbp]
\addtocounter{table}{-1}
\centering
\caption{(Continued.)}
\resizebox{\textwidth}{!}{%
\begin{tabular}{cccccccc}
\hline
&&&&&&&\\
Parent & Daughter & [$E_i, J^{\pi}_i$] & [$E_f, J^{\pi}_f$] & $Q_n$ (keV) & $\frac{\lambda_{bare}}{\lambda_{neutral}}|_{Par. Eqn.}$ & $\frac{\lambda_{bare}}{\lambda_{neutral}}|_{RF}$ & $\frac{\lambda_{bare}}{\lambda_{neutral}}|_{ANN}$\\
&&&&&&&\\
\hline
\hline

$^{184}$Ta & $^{184}$W & $[0,\;(5^-)]$ & $[1581.5,\;(6^-)]$ & 1284.5 & 1.22 & 1.17 & 1.12 \\

$^{184}$Ta & $^{184}$W & $[0,\;(5^-)]$ & $[1446.3,\;6^-]$ & 1419.7 & 1.20 & 1.16 & 1.10 \\

$^{184}$Ta & $^{184}$W & $[0,\;(5^-)]$ & $[1345.4,\;4^-]$ & 1520.6 & 1.18 & 1.16 & 1.08 \\

$^{184}$Ta & $^{184}$W & $[0,\;(5^-)]$ & $[1285,\;5^-]$ & 1581.0 & 1.18 & 1.16 & 1.07 \\

\hline
$^{187}$W & $^{187}$Re & $[0,\;3/2^-]$ & $[685.8,\;5/2^-]$ & 625.1 & 1.51 & 1.51 & 1.46 \\
\hline
\hline
\end{tabular}%
}
\end{table}

\section{\label{summary}Summary and Conclusion}

In this work, we employed artificial intelligence and machine learning (AIML) techniques to develop both a global parametric expression and predictive models based on the RF and ANN algorithms for estimating the ratio $(\lambda_{bare}/\lambda_{neutral})$ in fully ionised atoms. The derived parametrisation provides a computationally efficient and physically transparent framework for estimating $(\lambda_{bare}/\lambda_{neutral})$ over a broad range of nuclei.

The exponential form of the parametrisation reflects the strong sensitivity of bound state $\beta^-$ decay to the available decay phase space and the overlap of the electron wavefunction with the nucleus, both of which depend primarily on the decay energy $Q_n$ and the nuclear charge $Z$. The comparatively weak linear dependence on the mass number $A$ suggests that mass effects contribute mainly as higher order corrections, whereas the dominant behaviour is governed by $Z$ dependent terms. Furthermore, the observed evolution of the fitted coefficients with increasing $Z$, including the change in sign of the slope parameters and the emergence of quadratic and cubic dependencies, points to a complex interplay between nuclear structure effects and decay dynamics.

The machine learning models further demonstrate that the relationship between the governing nuclear parameters and the decay rate ratio can be learned with high accuracy. Among the investigated approaches, both the ANN and RF models exhibit exceptional predictive performance, outperforming the analytical parametrisation. This highlights the capability of unconstrained machine learning algorithms to capture the complex, nonlinear correlations inherent in bound state $\beta^-$ decay systematics.

Finally, the developed models were applied to a large set of nuclei relevant to stellar nucleosynthesis for which bare atom decay rate data are not currently available. The predicted decay rate ratios provide estimates of the maximum possible enhancement of $\beta^-$ decay under fully ionised conditions. These results provide useful estimation for astrophysical nucleosynthesis calculations, particularly in studies of abundance flow evolution, as well as for the planning and interpretation of future storage ring, plasma trap experiments involving highly charged ions.

\section*{Acknowledgments}

The authors acknowledge the IEDC Laboratory, Department of Basic Science and Humanities, IEM Kolkata (Newtown Sector), for providing computational facilities. The author AG gratefully acknowledges the guidance of Prof. Sukhendusekhar Sarkar and Dr. Chirashree Lahiri during earlier studies on the calculation of $\beta^-$ decay rates of bare atoms, which provided the foundation for the present work.

\noindent\textbf{Author Contributions:}
\textbf{Arkabrata Gupta:} Conceptualisation, computation, coding, data analysis and manuscript writing.
\textbf{Spandan Aich:} Computation, coding, and plotting.
\textbf{Suparna Sau:} Coding and manuscript writing.
\textbf{Sangeeta Das:} Data analysis and manuscript writing.

\section*{Data Availability}
The data that support the findings of this study are available upon reasonable request from the corresponding author. 


\bibliographystyle{unsrt}
\bibliography{Reference_jpg}

\end{document}